\documentclass[prx, reprint, amsmath, amssymb, longbibliography, superscriptaddress, floatfix, nofootinbib]{revtex4-2}

\usepackage[displaymath, mathlines]{lineno} 
\usepackage{etoolbox} 

\usepackage[utf8]{inputenc}
\usepackage{graphicx}

\usepackage[svgnames]{xcolor}
\usepackage{amsmath,amssymb,mathrsfs}
\usepackage{bm}
\usepackage[unicode=true, pdfusetitle, bookmarks=true, bookmarksnumbered=false, bookmarksopen=false, breaklinks=false, pdfborder={0 0 1}, backref=false, colorlinks=false]
 {hyperref}
\usepackage{caption}
\usepackage{subcaption}

\usepackage{ragged2e}
\DeclareCaptionJustification{justified}{\justifying}
\usepackage{diagbox}
\usepackage{braket,tensor,siunitx}

\usepackage[inkscapeformat=png]{svg}

\usepackage[percent]{overpic}
\usepackage{tikz}
\usepackage{array}
\usepackage{multirow}

\newcommand*\showwidth[1]{%
  \textcolor{blue}{\rule{\csname#1\endcsname}{1pt}}\newline
  \texttt{\textbackslash#1}: \expandafter\the\csname#1\endcsname
  \par
}

\makeatletter
\newcommand\thefontsize{The current font size is: \f@size pt}
\makeatother

\begin{document}
\title{Detection of weak signals under arbitrary noise distributions}

\author{J. Zschetzsche}
\thanks{These authors contributed equally to this work.}
\affiliation{Institute of Signal Processing, Johannes Kepler University Linz, Austria}
\affiliation{Christian Doppler Laboratory for Steel Industry Signal Processing and Machine Learning, Austria}

\author{M. Weimar}
\thanks{These authors contributed equally to this work.}
\affiliation{Institute for Theoretical Physics, Vienna University of Technology, Austria}

\author{O. Lang}
\affiliation{Institute of Signal Processing, Johannes Kepler University Linz, Austria}
\affiliation{Christian Doppler Laboratory for Steel Industry Signal Processing and Machine Learning, Austria}

\author{S. Schuster}
\affiliation{voestalpine Stahl GmbH, Linz, Austria}

\author{A. Haberl}
\affiliation{voestalpine Stahl GmbH, Linz, Austria}

\author{S. Schertler}
\affiliation{Institute of Signal Processing, Johannes Kepler University Linz, Austria}
\affiliation{Christian Doppler Laboratory for Steel Industry Signal Processing and Machine Learning, Austria}

\author{B. Lehner}
\affiliation{Silicon Austria Labs GmbH, Linz, Austria}

\author{J. Reisinger}
\affiliation{voestalpine Stahl GmbH, Linz, Austria}

\author{M. Huemer}
\affiliation{Institute of Signal Processing, Johannes Kepler University Linz, Austria}

\author{S. Rotter}
\affiliation{Institute for Theoretical Physics, Vienna University of Technology, Austria}

\begin{abstract}
Detecting weak signals buried in complex, non-Gaussian noise is a fundamental challenge in science and engineering, with applications ranging from radar systems and communications to industrial monitoring and gravitational wave detection.
The Rao detector, a key concept in this domain, achieves asymptotically optimal performance as the number of measurements increases, but requires precise knowledge of the data's statistical properties, often relying on simplified noise models.
We propose a hybrid framework that combines a lightweight neural network with the Rao detection framework to address this limitation.
The neural network, trained on noise-only data, learns the optimal multivariate nonlinearity, transforming noisy data to enhance signal detectability.
The newly introduced LRao detector then fully extracts the signal information, achieving asymptotically optimal performance even under challenging noise conditions.
Validated on both simulated and real-world magnetic sensor data, our method significantly outperforms conventional approaches.
By bridging data-driven techniques with model-based signal processing, it offers a robust and interpretable solution for signal detection across diverse applications.
\end{abstract}
\maketitle
\let\thefootnote\relax\footnotetext{Code and data are available under: \\\url{https://github.com/jonaslindenberger/LRao-detector}}

\section{Introduction}




When we try to follow a single conversation in a noisy environment – such as a crowded room – our brain performs a remarkable feat of signal detection: isolating a weak, structured signal from a complex acoustic background. 
This intuitive example captures the essence of a much broader challenge that spans science and engineering: detecting weak signals embedded in complex, often non-Gaussian noise environments. 
From radar \cite{richards2005fundamentals,coluccia2022adaptive,ramirez2023coherence} to nano-mechanical sensors \cite{bachtold2022mesoscopic} and gravitational wave detection \cite{martynov2016sensitivity}, the ability to discern faint signals from noisy data underpins progress in many fields. 
Yet, unlike the human auditory system, which can adapt its internal filters to varying acoustic conditions, most conventional detectors rely on simplified or idealized assumptions about the underlying noise statistics.

Overcoming this limitation lies at the core of modern detection theory. A common strategy here is to design experiments and data acquisition procedures that maximize the signal-to-noise ratio (SNR), thereby improving detection performance in subsequent processing. In many situations, however, the physical parameters that determine the SNR cannot be freely chosen. When a signal is buried in complicated, non-Gaussian noise, effective post-processing also becomes highly non-trivial. Such conditions occur, for example, when acoustic or electromagnetic waves are distorted by unpredictable scattering events, as in radar applications
 \cite{watts2022challenges,trunk2007detection,billingsley2002statistical}, naval measurements \cite{abraham2019underwater,abraham2011background} and complex light scattering \cite{starshynov2025model}. 
Other scenarios are vibration analysis \cite{jafarzadeh2025wave}, wireless sensor networks with quantized data \cite{mohammadi2022generalized,ciuonzo2017quantizer,wang2019detection}, and measurements of gravitational waves \cite{karnesis2025characterization,abbott2016observation,abbott2016characterization,yu2022nonlinear}.  
Non-Gaussian noise often manifests in signals from sensor data containing outliers. 
The conventional approach of removing these outliers would result in simpler statistics, but at the same time, one loses information about the signal.

An exact and widely used expression for the asymptotically optimal data-processing for the detection of a weak signal is provided by the so-called Rao score test or Rao detector \cite{rao1948large}.
The Rao detector can be interpreted \cite{kay1998fundamentals,lehmann1986testing} as applying a generally nonlinear transformation to the measured data such that the transformed data is maximally sensitive with respect to the signal of interest. 
This nonlinearity is the so-called score function or Fisher score, which is the gradient of the log-likelihood with respect to the signal parameters, and therefore indicates the sensitivity of the log-likelihood to parameter changes.

While the Rao detector is well-established for independent and identically distributed (IID) or Gaussian noise processes, its practical applicability breaks down when the probability distribution is unknown. 
When only a finite data set is available, one could, in the philosophy of frequentist statistics, estimate the corresponding probability density function (PDF) either with a parametric or a non-parametric approach and simply plug it into the analytical expression of the Rao detector. 
The issue with parametric approaches is that they typically require a very good model of the distribution. 
Non-parametric approaches, on the other hand, such as fitting a histogram to the data, are model-free but are hampered by the curse of dimensionality \cite{wasserman2006all}. 
Due to these issues, one typically assumes that the measurements are IID, resulting in a one-dimensional PDF and a scalar nonlinearity that accounts for the non-Gaussianity \cite{lehmann1986testing,kassam2012signal,kay1998fundamentals}.
To justify the assumption of independence, the data can be preprocessed by a whitening transformation, which is usually constrained to be linear.
However, this approach neglects nonlinear dependencies, which can contain a significant amount of information about the signal of interest. 
A simple example where the dependencies cannot be removed with a linear transformation is any signal corrupted by correlated noise with some additional outliers that do not show this correlation, e.g., due to sudden interferences or temporal sensor malfunctions \cite{zoubir2012robust,fox1972outliers,chen2022gross}.

\begin{figure*}
    \centering
    \includegraphics[width=1.\linewidth]{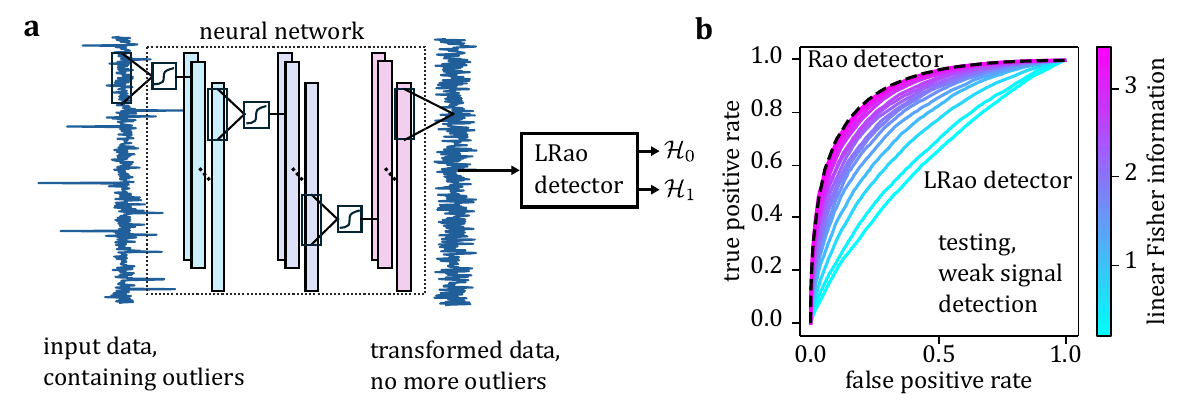}
    \caption{\textbf{Sketch of the appraoch and its performance.} \textbf{a}, A lightweight neural network is trained to optimally process the complicated, non-Gaussian data. By minimizing the cost function we introduce in this work, the neural network learns to represent the data in a way that is most suitable for further processing, no longer containing any outliers. When combined with an additive signal model, the neural network is trained exclusively on noise-only data, which is often available in abundance. \textbf{b}, As the cost function is minimized, the performance of the proposed LRao detector applied to the transformed data gradually improves until it reaches that of the asymptotically optimal Rao detector, which, in contrast to our method, requires precise knowledge of the data PDF. The cost function does not heuristically penalize outliers, instead, it arises naturally as the (negative) linear Fisher information. This quantity formally measures the information content in the transformed data that is easily extractable, without further complicated processing.}
    \label{fig:fig1}
\end{figure*}

In this work, we overcome all these difficulties by introducing a method for performing detection and estimation from data under complicated noise conditions in a data-driven manner, as illustrated in Fig.~\ref{fig:fig1}. 
Our key idea is to feed measurement data into a neural network, where the minimization of a suitable cost function improves the performance of a detector placed behind the network until it converges to the asymptotically optimal performance. 
This approach also comes with the benefit of interpretability as the cost function is directly related to the detector's performance in an intuitive and theoretically grounded way. 


We demonstrate the benefits of our new method on a problem from the field of weak signal detection. 
The problem is to detect a weak periodic signal from magnetic sensor data in the presence of approximately Gaussian ambient noise and non-Gaussian spiky geomagnetic noise.
We design a small convolutional neural network that we train purely on noise data (without any signal present) and show that our method's detection performance surpasses that of standard detectors by a large margin, converging to the asymptotically optimal performance.
Our neural network learns to effectively deal with outliers by transforming the heavy-tailed noise distribution to a short-tailed distribution, as illustrated in Fig.~\ref{fig:fig1}.
While the noise characteristics are learned from data, the signal knowledge is incorporated in a model-based fashion.

\section{theory of local parameter estimation and testing}\label{sec:theory}
Let us start by reviewing the seminal Rao detection statistic. 
Noisy measurement data is generally modeled by a random variable $\mathbf{x}\in\mathbb{R}^N$ following a probability distribution, which is determined by the PDF $\mathbf{x} \sim p(\mathbf{x};\boldsymbol{\theta})$, where $\boldsymbol\theta\in\mathbb{R}^l$ is a vector of $l$ unknown parameters.
The corresponding log-likelihood function is defined as $L(\mathbf{x},\boldsymbol\theta) = \log p(\mathbf{x};\boldsymbol\theta)$, where the arguments are dismissed in the following equations for the sake of readability.
A quantity that measures the information content within the data about a parameter of interest $\boldsymbol{\theta}$ is the so-called Fisher information (FI) \cite{fisher1922mathematical} defined as
\begin{equation}
\mathbf{F}=-\mathbb{E}\left[\nabla_{\boldsymbol{\theta}}\nabla_{\boldsymbol{\theta}}^\top L\right]
    \label{equ:definition_fim}\;,
\end{equation}
where $\mathbb{E}[\cdot]$ denotes expectation and $\nabla_{\boldsymbol\theta}=[\frac{\partial}{\partial[\boldsymbol\theta]_1}\dots\frac{\partial}{\partial[\boldsymbol\theta]_l}]$ is the nabla operator.
Being the expected local curvature of the likelihood function, the FI plays a pivotal role not only in parameter estimation but also in local parameter testing \cite{kay1993fundamentals,kay1998fundamentals,lehmann1986testing,scharf1991statistical}.
This becomes apparent as we discuss the Rao detector, which integrates the FI into its detection statistic.
The Rao detector decides between the so-called null hypothesis $\mathcal{H}_0$ and the alternative hypothesis $\mathcal{H}_1$ in a two-sided parameter test, defined as
\begin{eqnarray}
    &\mathcal{H}_0:\ \boldsymbol{\theta}=\boldsymbol{\theta}_0&\\
    &\mathcal{H}_1:\ \boldsymbol{\theta}\neq\boldsymbol{\theta}_0&.
    \label{equ:theory_hypo}
\end{eqnarray}
The parameter vector under $\mathcal{H}_1$, $\boldsymbol{\theta}_1$, is assumed to be ``close to'' the reference parameter vector $\boldsymbol{\theta}_0$ under $\mathcal{H}_0$.
The well-known Rao detection statistic is given by \cite{rao1948large}
\begin{equation}
    T_{\mathrm{Rao}}(\mathbf{x})=(\nabla_{\boldsymbol{\theta}}L_0)\mathbf{F}_0^{-1}(\nabla_{\boldsymbol{\theta}}L_0)^\top,
    \label{equ:Rao}
\end{equation}
where the subscript $0$ indicates that the expressions are evaluated at $\boldsymbol{\theta}_0$ (after differentiating). 
To obtain a decision about the underlying hypothesis, $T_{\mathrm{Rao}}$ is compared to a threshold that trades off the true positive rate (TPR) and the false positive rate (FPR).
Under $\mathcal{H}_0$, $T_\mathrm{Rao}$ asymptotically follows a central chi-squared $\chi^2_l$ distribution.
As a consequence, the FPR can be kept at a fixed value regardless of the unknown values of the parameter vector under $\mathcal{H}_1$.
Under $\mathcal{H}_1$, $T_{\mathrm{Rao}}$ is asymptotically non-central chi-squared $\chi'^2_l(\lambda_\mathrm{Rao})$ distributed with non-centrality parameter $\lambda_\mathrm{Rao}=(\boldsymbol\theta_{1}-\boldsymbol\theta_{0})^\top \mathbf{F}_0(\boldsymbol\theta_{1}-\boldsymbol\theta_{0})$.
As $\lambda_\mathrm{Rao}$ increases, the $\chi'^2_l(\lambda_\mathrm{Rao})$ distribution shifts towards higher values, and therefore the asymptotic TPR increases with the FI.
Although the Rao detector is asymptotically optimal under certain conditions \cite{lehmann1986testing}, its practical applicability is severely limited because the underlying PDF of the data is often unknown.
In such cases, the PDF is typically assumed in a way that enables the Rao detector to function as a robust and practical detector \cite{kay1998fundamentals}.

Our approach is to first make the bold assumption that the data follows a multivariate normal distribution $\mathcal{N}(\boldsymbol{\mu},\boldsymbol\Sigma_0)$, where $\boldsymbol{\mu}$ is the parameter dependent mean vector, and $\boldsymbol\Sigma_0$ is the covariance evaluated at the reference parameter $\boldsymbol{\theta}_0$.
This assumption is associated with the concept of an ``equivalent pessimistic system'', because the FI associated with this ``pessimistic PDF'' is the lowest among a broad class of data distributions \cite{stein2014lower,stein2015fisher,stoica2011gaussian}.
As the pessimistic PDF assumption is generally not valid, designing a detector based on it will produce suboptimal results.
However, in contrast to the conventional approach mentioned above, we now transform the data by feeding it into a neural network. 
The key is that we can still use the pessimistic PDF of the transformed data to construct a (suboptimal) detector. 
However, its performance is generally different from that of the original data. 
We are then able to set up an optimization problem: we select the parameters of the neural network such that for the transformed data, the pessimistic PDF assumption leads to statistically optimal results. 
In the following, we introduce the performance metric that we use as the optimization objective for tuning the neural networks' weights. 
As it turns out, there is a natural choice for this objective, which is determined by the so-called linear Fisher information (LFI) \cite{series2004tuning,moreno2014information,kanitscheider2015measuring,kohn2016correlations,cheng2023representational,weiss2024measuring,zylberberg2016direction,hu2014sign}.

The LFI corresponds exactly to the FI calculated under the pessimistic PDF assumption and provides a lower bound for the FI associated with the original PDF \cite{stein2014lower}. 
Evaluated at $\boldsymbol{\theta}_0$, the LFI is given by 
\begin{equation}
    \mathbf{J}_0=(\nabla_{\boldsymbol{\theta}}\boldsymbol{\mu}_0)^\top \boldsymbol{\Sigma}_0^{-1}(\nabla_{\boldsymbol{\theta}}\boldsymbol{\mu}_0).
    \label{equ:LFI}
\end{equation} 
Its inverse provides an attainable lower bound for the covariance matrix of a locally unbiased linear estimator \cite{kohn2016correlations}.
The locally best linear unbiased estimator (LBLUE), whose covariance matrix equals the inverse LFI and hence attains the bound, is given by
\begin{equation}
    \hat{\boldsymbol\theta}_{\mathrm{LBLUE}}=\boldsymbol\theta_0+\mathbf{J}^{-1}_0(\nabla_{\boldsymbol{\theta}}\boldsymbol{\mu}_0)^\top \boldsymbol{\Sigma}_0^{-1}(\mathbf{x}-\boldsymbol\mu_0).
    \label{equ:lblue}
\end{equation}
Under certain conditions its asymptotic distribution is given by $\mathcal{N}(\boldsymbol\theta,\mathbf{J}_0^{-1})$
as proven in Supplementary Information (SI)~\ref{app:theory_LBLUE}.
To the best of our knowledge, we present here the multivariate form of the LBLUE for the first time. 
In the special case of a scalar parameter, the LBLUE was first derived in \cite{moreno2014information}, and is also referred to as the locally optimal linear estimator, or equivalently, the locally optimal linear decoder \cite{kanitscheider2015measuring,kohn2016correlations}.

We now extend the concept of local unbiased estimation in the LFI framework to local parameter testing or detection.
By evaluating the Rao detection statistic \eqref{equ:Rao} under the pessimistic PDF assumption, we propose a ``pessimistic Rao detector'', which we call the linear Rao (LRao) detector.
As shown in SI~\ref{app:theory_LRao}, the LRao detection statistic is given by
\begin{align}
    &T_\mathrm{LRao}(\mathbf{x})=\nonumber\\
    &(\mathbf{x}-\boldsymbol\mu_0)^\top \boldsymbol\Sigma_0^{-1}(\nabla_{\boldsymbol\theta}\boldsymbol\mu_0)\mathbf{J}^{-1}_{0}(\nabla_{\boldsymbol\theta}\boldsymbol\mu_0)^\top \boldsymbol\Sigma_0^{-1}(\mathbf{x}-\boldsymbol\mu_0).
\end{align}
Although it is quadratic in the data, we include linear in its name to be consistent with the LFI and the LBLUE, and to emphasize that the pessimistic score function, which lies at the heart of the LRao detector, is actually linear, as discussed in SI~\ref{app:theory_pess}.
Similar to the Rao detector, under mild assumptions the LRao detection statistic asymptotically follows a central $\chi_l^2$ distribution under $\mathcal{H}_0$, and a $\chi_l'^2(\lambda_\mathrm{LRao})$ distribution with non-centrality parameter $\lambda_\mathrm{LRao}=(\boldsymbol\theta_{1}-\boldsymbol\theta_{0})^\top \mathbf{J}_0(\boldsymbol\theta_{1}-\boldsymbol\theta_{0})$ under $\mathcal{H}_1$.
Therefore, its TPR increases as the LFI increases and matches that of the Rao detector if the LFI equals the FI.

A different perspective on the LRao detector can be obtained by expressing the detection statistic in terms of the LBLUE and the LFI as
\begin{equation}
    T_{\mathrm{LRao}}=(\hat{\boldsymbol\theta}_{\mathrm{LBLUE}}-\boldsymbol\theta_0)^\top \mathbf{J}_{0}(\hat{\boldsymbol\theta}_{\mathrm{LBLUE}}-\boldsymbol\theta_0).
\end{equation}
The LRao detector can thus be interpreted as first estimating the parameter vector using the LBLUE, and then normalizing the squared deviation from the reference parameter by its covariance, i.e., the inverse LFI.
This yields a normalized squared distance measure, which is used to detect a change in the parameter vector.
Note that the squaring operation is necessary due to the two-sided nature of the hypothesis test in \eqref{equ:theory_hypo}.
For a one-sided parameter test, where the alternative hypothesis $\mathcal{H}_1$ is defined as $\theta>\theta_0$ (and the parameter is typically a scalar signal strength measure \cite{kassam2012signal,song2002advanced,kay1998fundamentals}), no squaring operation is required.
In that case, the (normalized) LBLUE itself can be used as a locally optimal linear detector, which we call the linear locally most powerful (LLMP) detector, see SI~\ref{app:theory_LLMP} for details.
As noted above, while the scalar LBLUE was first derived in \cite{moreno2014information}, the LLMP detector is, to the best of our knowledge, presented here for the first time in the context of detection theory.

We now discuss the transformation that causes the LBLUE (therefore also the LLMP detector) and the LRao detector to be asymptotically optimal.
Nonlinear transformations, such as those performed by a neural network, can modify the LFI \cite{weimar2025fisher}, and the network can be optimized to maximize the LFI at its output. 
Since, however, the LFI is a lower bound for the FI, maximizing the LFI effectively yields the FI, provided the network can learn the optimal transformation.
The LFI can be estimated by substituting the mean and covariance matrix in \eqref{equ:LFI} with sample estimates and approximating derivatives using central difference quotients \cite{friedlander2022data}.
For a scalar parameter, the estimated LFI at the output of a neural network can be directly used for optimization.
For a vector parameter, however, one cannot directly use the estimated LFI since a scalar cost function is required.
We found the trace of the estimated LFI to be particularly effective, as it is maximized when the data is transformed by the score function (or an invertible affine transform of the score function), as shown in SI~\ref{app:theory_LFImaxscore}.
If such a transformation is applied to the data as a pre-processing step, local estimation and detection reduce to problems, which are asymptotically optimally solved by the LBLUE and the LRao detector, respectively.
Thus, by maximizing the trace of the estimated LFI $\widehat{\mathbf{J}}_0$ at the output of a neural network, or equivalently by minimizing the cost function
\begin{equation}
    C=-\mathrm{tr}(\widehat{\mathbf{J}}_0),
    \label{equ:lfi_cost_fun}
\end{equation}
we effectively learn the score function from the data, bypassing the need for probability density estimation.


\begin{figure*}
    \centering
    \includegraphics[width=1\linewidth]{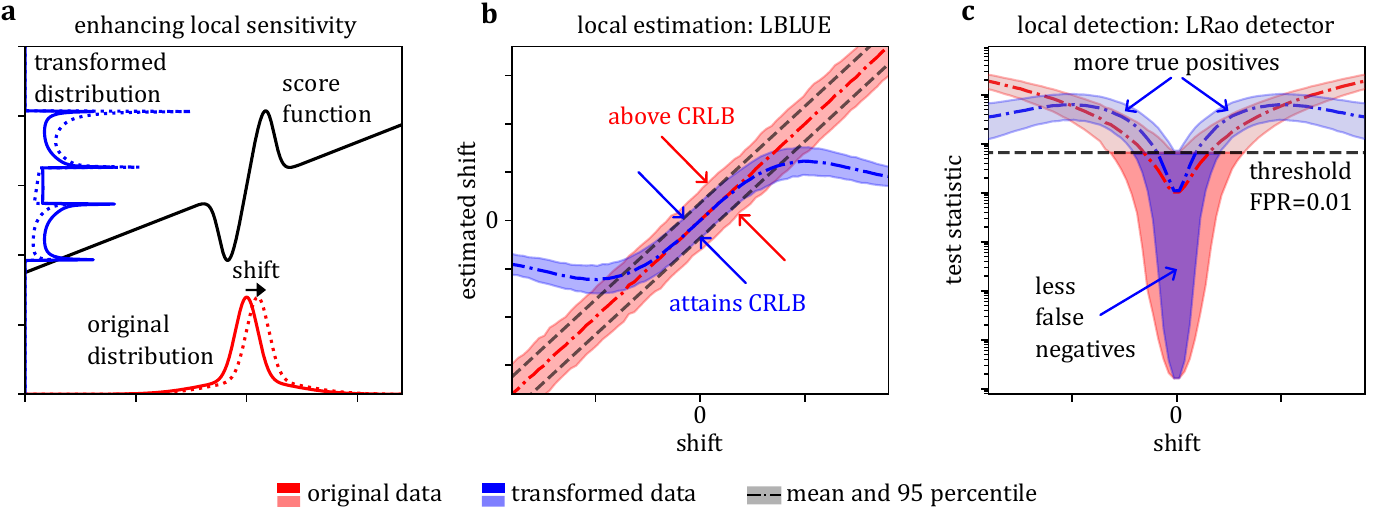}
    \caption{\textbf{Noise samples drawn from a mixture of Gaussians.} \textbf{a}, In our task to estimate and detect a shift in a vector of IID noise samples, the original distribution (mix auf Gaussiens, red) is transformed to a new distribution (blue) using the score function (black), thereby maximizing the LFI that the LBLUE can extract. \textbf{b}, Means and the 95 percentiles of the LBLUE estimates for both the original and the transformed data. Notably, the LBLUE applied to the transformed data attains the CRLB locally, but yields biased estimates when considered globally. The LBLUE applied to the original data is unbiased globally, however, its variance lies above the CRLB. \textbf{c}, LRao detection statistics and threshold. Here, the LRao detector applied to the score-transformed data, which is equivalent to the Rao detector on the original data, demonstrates a higher local probability of detection compared to the LRao detector on the original data. This can be seen by comparing the color-filled areas between the blue and the red curves, respectively, above and below the horizontal dashed line representing the threshold.}
    \label{fig:fig2}
\end{figure*}
We now present a simulated example to demonstrate how transforming data with the score function enhances performance in local parameter estimation and detection using the LBLUE and the LRao detector.
The aim is to estimate and detect an unknown additive shift parameter in a vector of IID noise samples originating from a Gaussian mixture with identical means but differing variances.
Instead of using a neural network for this example, we employ the optimal nonlinear transformation, i.e., the score function, which is readily available.
In Fig.~\ref{fig:fig2}a, the original data distribution (red) and the distribution of the data transformed by the score function (blue) are shown.
The tails of the original distribution are scaled down, and both Gaussians in the mixture are approximately linearly mapped from their dominant regions into the same interval.
In Fig.~\ref{fig:fig2}b, the means and the 95 percentiles of the LBLUE for the original and the score-transformed data are shown.
Applying the score function causes the LBLUE to locally attain the Cramér-Rao lower bound (CRLB) (i.e., the inverse FI), as the LBLUE can locally extract all the information about the shift parameter, without the need for another nonlinearity.
In Fig.~\ref{fig:fig2}c, the means and the 95 percentiles of the LRao detection statistics are shown.
Combined with the score function, the LRao detector becomes equivalent to the Rao detector and is asymptotically optimal.

At this point, we emphasize that for the task of weak signal detection, locality is not a rough approximation but exactly what one can expect for the following reasons:
First, detecting a signal is most difficult if the SNR is small, i.e., the signal is weak.
If the SNR is high, detection is comparably simple, as one can simply settle for suboptimal solutions.
Second, many practically relevant score functions will produce reliable results for large data records, even if the signal is not weak.
As can be seen in Fig.~\ref{fig:fig2}c, the original data produces a larger statistic than the transformed data for large shifts, however, the transformed data still lies above the threshold.
Practical score functions for the task of signal detection typically resemble some kind of ``limiter function'' \cite{kay1998fundamentals}, causing the Rao detection statistic to exceed a reasonably set threshold if the data vector size is large.
Notably, large data vector sizes are essential for reliable detection performance, especially in low-SNR environments. 

We emphasize that our LFI optimization framework is compatible with the LBLUE for local estimation, the LRao detector for local two-sided detection, and the LLMP (normalized LBLUE) for local one-sided detection. 
In all three cases, increasing the LFI improves performance, with each method ultimately approaching its respective asymptotically optimal limit.
In the remainder of this work, we focus on the LRao detector for the task of detecting a weak signal in additive stationary noise.

\section{weak signal detection}\label{sec:weak}
We now apply the proposed theory to weak signal detection, demonstrating our data-driven strategy for learning the optimal multivariate nonlinearity directly from measurements.
We employ a neural network to parameterize this nonlinearity and select its weights by maximizing the trace of the estimated LFI.
The assumptions of additivity, enabling the use of noise-only data for training, and stationarity, which enhances the efficiency and robustness of the training process, are discussed.
Detailed derivations can be found in SI~\ref{app:weak}.

One of the most important practical applications of local parameter testing is weak signal detection in non-Gaussian noise \cite{kay1998fundamentals,song2002advanced,kassam2012signal,sheehy1978optimum}.
For the two-sided parameter test described above, the local parameter vector $\boldsymbol{\theta}$ typically represents the amplitudes of the signal components to be detected, with $\boldsymbol{\theta}_0=\mathbf{0}$ as the local operating point.
The simplest and most widely used model is the linear signal model in additive noise, defined as
\begin{equation}
    \mathbf{x}=\mathbf{H}\boldsymbol{\theta}+\mathbf{w},
    \label{equ:additive_signal_model}
\end{equation}
where $\mathbf{H}\in\mathbb{R}^{N\times l}$ is the known observation matrix, whose columns span the signal space, and $\mathbf{w}\in\mathbb{R}^{N}$ is the noise vector following an unknown distribution.
As detailed in SI~\ref{app:weak_additive}, the additive model \eqref{equ:additive_signal_model} offers the advantage that the Jacobian of the mean (after transforming the data with a neural network) can be computed exclusively from noise-only ($\mathcal{H}_0$) data, which is often available in abundance. 
The proposed method extends to cases where the observation matrix $\mathbf{H}$ is (partially) unknown during training. 
This situation arises, for instance, when a nuisance parameter exists only under the alternative hypothesis $\mathcal{H}_1$, as in the detection of a periodic signal with an unknown period length \cite{davies1987hypothesis,kay2013fundamentals}. 
In such cases, the optimization can be reformulated for the model $\mathbf{x}=\boldsymbol\theta^*+\mathbf{w}$ by applying the chain rule, with the Jacobian of the mean post-multiplied by the observation matrix, as explained in SI~\ref{app:additive_noise}.

Constructing the optimal detector by maximizing the LFI becomes significantly simpler when the noise arises from a stationary process, where statistical properties remain constant over time. 
This assumption, common for time series data, is frequently valid in practical applications and enables the use of powerful analytical tools \cite{papoulis2002probability}.
In this case, the covariance matrix of the data is a symmetric Toeplitz matrix and its inverse can be approximated as \cite{gray2006toeplitz,kay1998fundamentals}
\begin{equation}
    \boldsymbol{\Sigma}^{-1}\approx\sum\limits_{i=0}^{N-1}\frac{\mathbf{v}_i\mathbf{v}_i^H}{P_{xx}(f_i)}\;,
    \label{equ:inverse_covariance}
\end{equation}
where $H$ denotes the conjugate transpose, $\mathbf{v}_i^H=\frac{1}{\sqrt{N}}\begin{bmatrix}1 & \mathrm{e}^{-\mathrm{j}2\pi f_i} & \mathrm{e}^{-\mathrm{j}4\pi f_i} & \hdots & \mathrm{e}^{-\mathrm{j}2\pi (N-1)f_i}\end{bmatrix}$ is the $i$th discrete Fourier transform basis vector with normalized frequency $f_i=i/N$ for $i=0,1,...,N-1$, and $P_{xx}(f_i)$ is the power spectral density of the data. 
Using \eqref{equ:inverse_covariance} simplifies the evaluation of the LFI \eqref{equ:LFI} in two ways. 
First, it enables direct evaluation of the inverse covariance matrix, eliminating the need for numerical inversion of the sample covariance matrix.
Second, it drastically reduces the amount of parameters that need to be estimated.
This approach is described in detail in SI~\ref{app:weak_stationary}.
Combining \eqref{equ:inverse_covariance} with \eqref{equ:lfi_cost_fun} and \eqref{equ:LFI}, the neural network $\Psi_W(\mathbf{x})$, parameterized by a set of learnable parameters $W$, can be optimized by minimizing
\begin{equation}
    C(\Psi_W(\mathbf{x}))= - \sum\limits_{j=1}^{l}\sum\limits_{i=0}^{N_{\Psi}-1}\frac{|\mathbf{v}_i^H\frac{\partial\boldsymbol{\mu}_\Psi}{\partial\theta_j}|^2}{{P_{\Psi \Psi}(f_i)}}
    \label{equ_cost_function}
\end{equation}
where $\boldsymbol{\mu}_{\Psi}$ is the mean of the output data $\Psi(\mathbf{x})$ (the subscript $W$ is omitted to keep the notation uncluttered), $P_{\Psi \Psi}(f_i)$ the power spectral density of $\Psi(\mathbf{x})$, $N_{\Psi}$ the dimension of $\Psi(\mathbf{x})$ (which we set equal to the input dimension $N$), and $|\cdot|$ the modulus of a complex number. 
During the optimization, \eqref{equ_cost_function} is evaluated by approximating $\boldsymbol{\mu}_{\Psi}$ by batch averages, $\partial\boldsymbol{\mu}_\Psi/\partial\theta_j$ by a central difference quotient, and $P_{\Psi \Psi}(f_i)$ by the batch-averaged periodogram \cite{schuster1898investigation,bartlett1948smoothing}.
Intuitively, the cost function is the trace of the LFI evaluated in the DFT domain.
The simple form is because the eigenvectors of the covariance matrix $\boldsymbol\Sigma_{\Psi\Psi}$ asymptotically align with the DFT vectors, and the product of the Jacobian of the mean with the covariance matrix, see \eqref{equ:LFI}, yields a squared DFT of the Jacobian's columns.
With the assumptions of additivity and stationarity, we can efficiently optimize our lightweight neural network using only a small amount of data, as demonstrated in the real-world example below.

We now specify the architecture of our neural network.
Our model leverages the time-translational symmetry of stationary data by employing convolutional layers that act upon the data as filters, only taking into account data points in a small range of time steps, as shown schematically in Fig.~\ref{fig:fig1} and in more detail in SI~\ref{app:weak_neural}. 
Additionally, we assume the noise distribution is symmetric about zero, resulting in an odd score function. 
This property is embedded in the network design by using tanh activation functions and omitting bias parameters in the convolutional layers.
To ensure scale invariance, the input data is normalized using a robust estimate of the signal scale, thereby enforcing invariance through invariant inputs \cite{coluccia2020cfar}.

While this architecture involves only a small number of trainable parameters, we show in SI~\ref{app:weak_neural} that it does not suffer a loss of performance compared to larger models. 
Combined with our computationally cheap cost function \eqref{equ_cost_function}, this neural network model enables fast training, making it suitable for real-time applications and online adaptive strategies. 

\begin{figure*}[!t]
    \centering
    \includegraphics[width=1\linewidth]{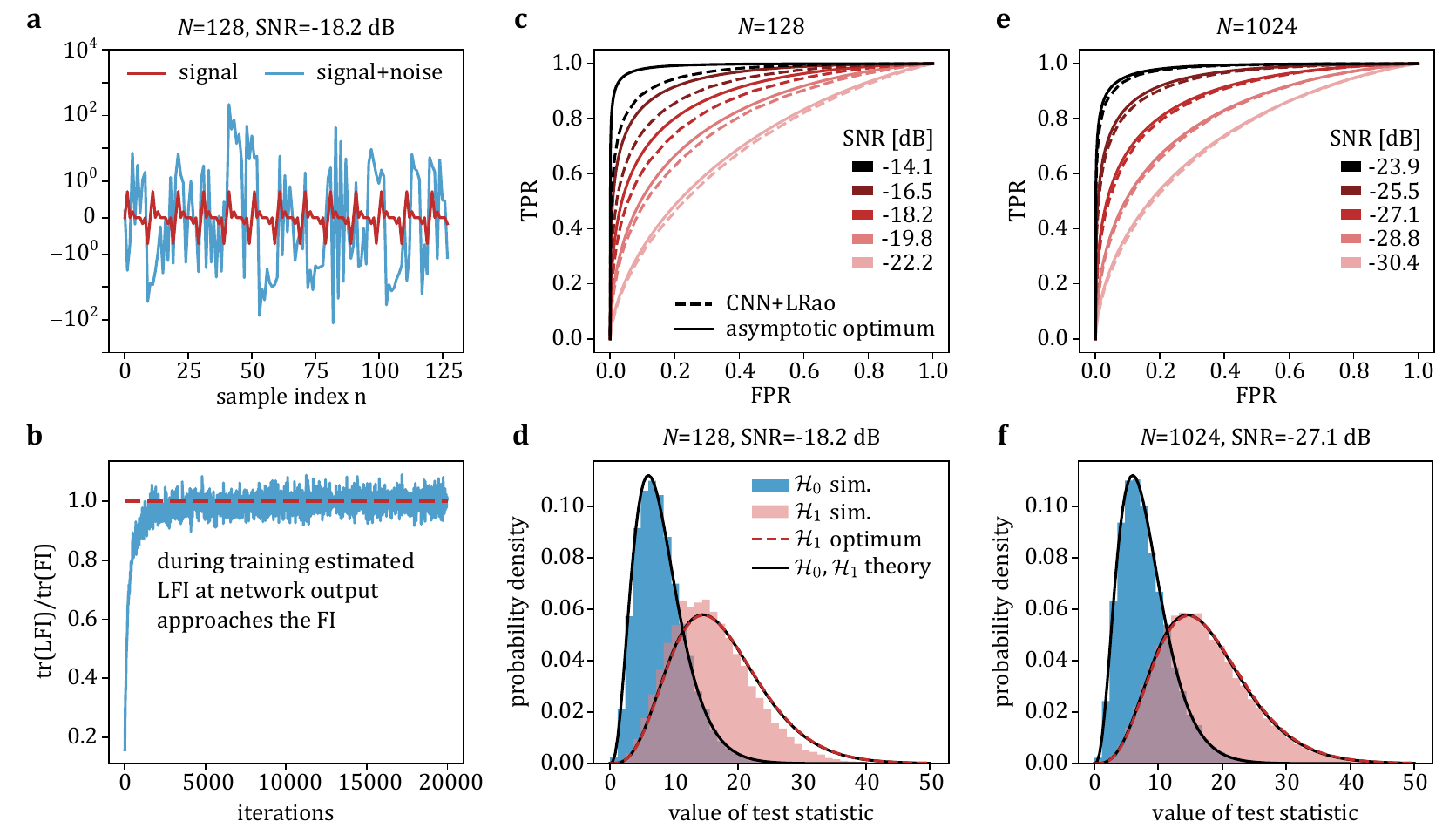}
    \caption{\textbf{Simulated Cauchy noise.} \textbf{a}, The periodic signal to detect and the signal corrupted by additive heavy-tailed Cauchy noise. For improved visibility, the scale of the ordinate is linear within the interval $[-1,1]$ and logarithmic outside of the interval. \textbf{b}, The trace of the LFI, estimated at the output of the neural network, normalized by the trace of the FI. \textbf{c}, The simulated and the asymptotically optimal ROC curves for different SNRs for a sequence length of $N=128$. The LRao detector does not yet attain its asymptotic performance. \textbf{d}, The simulated and analytical distributions of the LRao detector. Under $\mathcal{H}_1$, the LRao detection statistic is slightly shifted to the left compared to the analytical asymptotic distribution. \textbf{e}, The ROC curves for $N=1024$. The LRao detector approaches its asymptotic performance, as can also be seen in \textbf{f}.}
    \label{fig:cauchy}
\end{figure*}
\subsection{Simulation: Cauchy noise}\label{subsec:cauchy}
First, we verify our method on simulated noise data. 
We sample data from a distribution where the FI and, therefore, the optimal value of our cost function is known, which provides us with a means of objective comparison.
Specifically, we opted for Cauchy noise to demonstrate that our method is effective even for this heavy-tailed distribution with undefined higher-order moments.
To generate noise sequences, we sample IID vectors of Cauchy noise and subsequently apply a linear filter to introduce dependency between the samples, enabling closed-form computation of the FI for the additive noise model~\eqref{equ:additive_signal_model}, as discussed in SI~\ref{app:cauchy}.
Similar to the real-world example discussed below, we are designing a detector for weak periodic signals of known period, which are modeled as a Fourier series and can be cast into the signal model~\eqref{equ:additive_signal_model}, as shown in SI~\ref{app:weak_periodic}.

In Fig.~\ref{fig:cauchy}a, the signal to detect and the noise-corrupted signal are shown.
Due to the heavy tails of the Cauchy distribution, the scale of the ordinate is linear within the interval $[-1,1]$ and logarithmic outside of the interval.
Fig.~\ref{fig:cauchy}b illustrates that, during training, the neural network increases the trace of the estimated LFI at its output until it approaches the trace of the FI.
As the LFI equals the FI, the LRao detector coincides with the Rao detector and is therefore asymptotically optimal.
Fig.~\ref{fig:cauchy}c shows the receiver operating characteristic (ROC) curves of the LRao detector and the asymptotically optimal ROC curves for a sequence length of $N=128$ at different SNRs.
The LRao detector does not yet attain its asymptotic performance, as can also be seen in Fig.~\ref{fig:cauchy}d.
The histogram under $\mathcal{H}_1$ is slightly shifted to the left compared to the asymptotic distribution, indicating that, for this example, the weak signal assumption is already slightly inaccurate at this SNR level.
In contrast, for the real-world example presented below, it will be seen that the LRao detector already attains its asymptotic performance for $N=128$ in good approximation.
Fig.~\ref{fig:cauchy}e shows the ROC curves for an increased sequence length of $N=1024$.
The LRao detector now almost matches the asymptotically optimal performance, which can also be deduced from the distributions shown in Fig.~\ref{fig:cauchy}f.
Note that the same lightweight convolutional neural network was used (without retraining) for the different sequence lengths, as the optimal nonlinear filter operation performed by the network is independent of the sequence length due to the stationarity of the noise samples.

\begin{figure*}[!t]
    \centering
    \includegraphics[width=1\linewidth]{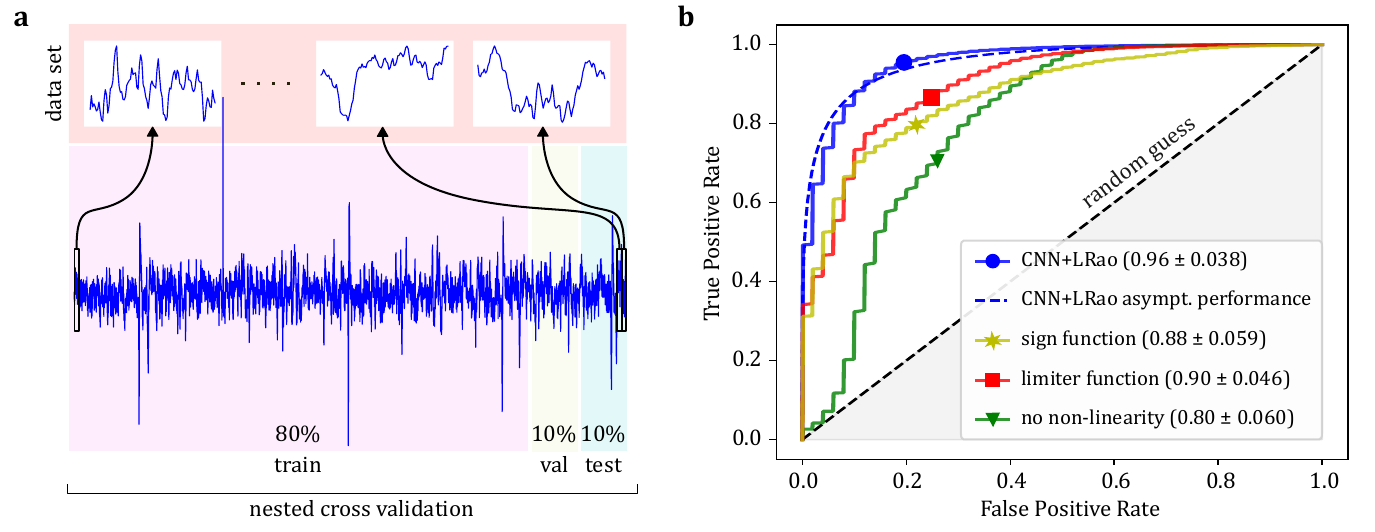}
    \caption{\textbf{Experimental data from magnetic sensors.} \textbf{a}, The spikey and dependent noise data captured by a magnetic sensor, available in the supplementary material of \cite{kay2013fundamentals}. The data sequence of length \num{10000} is split into smaller sequences of length \num{128}, forming a data set of \num{78} sequences. A nested cross-validation procedure is used to evaluate the performance of our detector, which aims to detect a weak periodic signal that is added in post-processing. \textbf{b}, ROC curves and areas under the curves with 95 percentiles (given in the brackets) for an input SNR of \SI{-25.5}{\decibel}. The proposed LRao detector in combination with the lightweight convolutional neural network (CNN+LRao) clearly outperforms the reference methods. The CNN+LRao detector attains its asymptotic performance in good approximation, and is therefore optimal to the degree that the LFI at the network output matches the original data's FI. We emphasize the effectiveness of our approach despite the limited amount of training data.}
    \label{fig:data_set}
\end{figure*}
\subsection{Real-world example: Magnetic sensor data}

We now demonstrate the effectiveness of our approach using real-world measurement data, as depicted in Fig.~\ref{fig:data_set}a. 
This data, collected by a magnetic sensor, comprises approximately Gaussian ambient noise alongside spiky geomagnetic noise. 
A detailed description of the data acquisition process can be found in \cite[p.545 ff.]{kay2013fundamentals}, as the data is taken from the supplementary material of this book.
The data sequence is segmented into smaller, non-overlapping sequences to create a dataset, which is subsequently utilized for training, validation, and testing through an 8-1-1 nested cross-validation procedure.
The task is to design a detector tailored to weak, periodic signals of known period length based on the noise-only measurement data at hand.
As in the simulation example in Subsection~\ref{subsec:cauchy}, the periodic signal is modeled as a Fourier series, which can be cast into the model \eqref{equ:additive_signal_model}, as discussed in SI~\ref{app:weak_periodic}.

As we transition to real-world data, we compare our method against standard reference detectors. 
These reference detectors follow a series of processing steps: a robust linear whitening transformation to mitigate dependencies in the noise data and justify the assumption of independence, a scalar function to handle outliers, and the generalized likelihood ratio test (GLRT) statistic under the assumption of IID Gaussian noise (see SI~\ref{app:weak_ref}).
It is important to note that for IID non-Gaussian noise, computing the GLRT statistic under the assumption of IID Gaussian noise for score-transformed data results in the Rao detector, which is asymptotically optimal \cite{kay1998fundamentals}. 
Consequently, the reference methods can be viewed as Rao detectors with manually designed score functions.
Specifically we use the sign function, which equals the score function for the Laplace distribution (used in the text book \cite{kay2013fundamentals} from where the data is taken), the (heuristic) limiter function, which cuts of values larger than three times the standard deviation, and the identity function, which equals the score function for the standard Gaussian distribution.

Although certain segments of the data are reasonably approximated by a Gaussian distribution (see SI~\ref{app:weak_magnetic}), we anticipate that the Gaussian assumption, i.e., the absence of nonlinearity, will yield suboptimal results compared to robust methods, as even a single outlier can trigger a false positive.
In alignment with the training, validation, and testing of the neural network, the reference detectors are assessed through a cross-validation procedure employing a 9-1 train-test split. 
Since no hyperparameters are optimized, a validation set is unnecessary. 
The whitening matrix is derived from the training data, and the performance is subsequently evaluated on the testing data.

In Fig.~\ref{fig:data_set}b, the ROC curves of the proposed detector (CNN+LRao, blue), its asymptotic performance, and the reference detectors are compared.
The asymptotic performance is computed using the LFI estimated from the training plus validation data, averaged over all cross-validation runs.
Our CNN+LRao detector attains its asymptotic performance in good approximation and is therefore optimal to the degree that the LFI at the output of the CNN matches the original data's FI.
Our proposed detector outperforms all the reference detectors by a large margin, including the ``sign detector'' used in \cite{kay2013fundamentals}, despite the limited amount of training data available.
The poor performance of the Gaussian GLRT detector (no nonlinearity) highlights the necessity of dealing with heavy-tailed noise.
Our method inherently penalizes these heavy tails and learns the optimal multivariate nonlinearity directly from noise-only measurement data, rather than relying on heuristic limiter designs or assuming a specific noise probability distribution.
\section{Summary and Outlook}
Our approach, centered on maximizing the linear Fisher information of a dataset, enables the detection of signals buried in strong, dependent, non-Gaussian noise. 
By solving the optimization problem, we achieve a transformation of the noisy data that effectively addresses non-Gaussianities, thereby making all the information in the data readily extractable by the LBLUE and the LRao detector.
This transformation is parameterized using a lightweight neural network architecture, specifically designed for processing stationary time-series data. 
Once the network is well-trained, we apply the LRao test to the transformed data, which is equivalent to performing the Rao test on the original data.
We demonstrate that our method converges to the optimal solution for signals buried in simulated Cauchy noise and significantly outperforms conventional approaches when applied to experimental data from magnetic sensor measurements, even with a limited amount of training data. 
Unlike other methods, our approach is asymptotically optimal and does not require explicit knowledge of the underlying probability distribution, as the optimal multivariate nonlinearity is directly learned from noise-only data.

Our approach offers a promising advancement on the extensive body of work concerning local nonlinear detection statistics \cite{kassam2012signal,song2002advanced,lehmann1986testing,kay1998fundamentals}, providing a powerful and computationally efficient alternative to model-based methods that often necessitate detailed examination and statistical modeling of the data. 
Unlike conventional deep learning strategies, it does not need to serve as a data-driven black box (although it can) but allows for hybrid usage, since model knowledge, which is frequently both available and accurate in engineering systems, can be easily integrated. 
Although this work focuses on weak signals in additive noise, the approach can be readily adapted to accommodate also other signal models.

A remarkable property of our method that sets it apart from other techniques is that it is well-suited for extension to online adaptive strategies. 
Due to its small size, the neural network does not require large amounts of data, allowing for high sensitivity to new measurements. 
This is of great importance in many detection tasks, for example, in physical experiments or industrial processes where drifts of the setup and external perturbations need to be taken into account by continual monitoring.

\section*{Acknowledgements}
We are grateful to Lukas M. Rachbauer for stimulating discussions and constructive suggestions that helped shape this study. 
The financial support by the BMWET, the National Foundation for Research, Technology and Development, and the Christian Doppler Research Association is gratefully acknowledged. 
This work was supported by Silicon Austria Labs (SAL), owned by the Republic of Austria, the Styrian Business Promotion Agency (SFG), the federal state of Carinthia, the Upper Austrian Research (UAR), and the Austrian Association for the Electric and Electronics Industry (FEEI).
\clearpage
\onecolumngrid
\renewcommand{\appendixname}{Supplementary Information}
\appendix
\section{Theory}\label{app:theory}
The following provides more detailed derivations of the equations presented in Section \ref{sec:theory}.
Additionally, we provide some background information on the topic of local estimation and detection.

\subsection{Local likelihood expansion}\label{app:theory_LMLE}
In this work, we are interested in local parameter estimation and local hypothesis testing, e.g., weak signal detection.
By local, we mean that the parameter vector $\boldsymbol\theta$ is assumed to be close to a reference parameter vector $\boldsymbol\theta_0$.
In this case, we can make use of the second-order Taylor series expansion of the log-likelihood function $L=\log p(\mathbf{x};\boldsymbol\theta)$ around a reference parameter $\boldsymbol{\theta}_0$
\begin{equation}
    L\approx L_0+\nabla_{\boldsymbol\theta}L_0(\boldsymbol\theta-\boldsymbol\theta_0)
    +\frac{1}{2}(\boldsymbol\theta-\boldsymbol\theta_0)^\top \nabla_{\boldsymbol{\theta}}\nabla_{\boldsymbol{\theta}}^\top L_0(\boldsymbol\theta-\boldsymbol\theta_0),
\end{equation}
where the subscript $0$ denotes that the expressions are evaluated at $\boldsymbol{\theta}=\boldsymbol{\theta}_0$ after the gradient and the Hessians are applied. 
Under very general circumstances \cite{bar1971asymptotic}, the quantity $\nabla_{\boldsymbol{\theta}}\nabla_{\boldsymbol{\theta}}^\top  L_0$ can be replaced by its expectation, the negative FI, which yields
\begin{equation}
    L\approx L_0+\nabla_{\boldsymbol\theta}L_0(\boldsymbol\theta-\boldsymbol\theta_0)
    -\frac{1}{2}(\boldsymbol\theta-\boldsymbol\theta_0)^\top \mathbf{F}_0(\boldsymbol\theta-\boldsymbol\theta_0).
    \label{equ:appTSELL}
\end{equation}
This local expression for the log-likelihood provides valuable insight into the FI and its role in local parameter estimation and detection.

We start by discussing the local maximum likelihood estimator (LMLE), which is very reminiscent of the LBLUE and naturally leads us to the Rao detection statistic.
Generally, the MLE for $\boldsymbol{\theta}$ satisfies $\nabla_{\boldsymbol\theta}L|_{\boldsymbol\theta_{\mathrm{MLE}}}=\mathbf{0}^\top $.
Taking the derivative of \eqref{equ:appTSELL} with respect to $\boldsymbol\theta$ yields an approximation for the score function
\begin{equation}
    \nabla_{\boldsymbol\theta}L\approx \nabla_{\boldsymbol\theta}L_0
    -(\boldsymbol\theta-\boldsymbol\theta_0)^\top \mathbf{F}_0,
    \label{equ:app_local_score}
\end{equation}
which is a linear function of $\boldsymbol\theta$.
Setting to zero and solving for $\boldsymbol{\theta}$ leads to the LMLE
\begin{equation}
    \hat{\boldsymbol\theta}_{\mathrm{LMLE}}=\boldsymbol\theta_0+\mathbf{F}^{-1}_{0}(\nabla_{\boldsymbol\theta}L_0)^\top .
    \label{equ:appLMLE}
\end{equation}
This estimator is discussed in \cite{alsing2018generalized} under the name of quasi-MLE for the purpose of optimal data compression.
This is due to the fact that the score function, or equivalently the LMLE in \eqref{equ:appLMLE}, is a local sufficient statistic for the parameter vector $\boldsymbol\theta$ since $\nabla_{\boldsymbol\theta}L_0$ is the only data-dependent term.
The score function reduces the dimensionality of the data from $N\rightarrow l$, while keeping the FI in the data constant.
When applied iteratively, \eqref{equ:appLMLE} is commonly referred to as scoring, or Fisher's scoring method \cite[p.187]{kay1993fundamentals}.
The scoring method converges to the maximum likelihood estimate (MLE) if the MLE approaches its large sample distribution $\mathcal{N}(\boldsymbol{\theta},\mathbf{F}^{-1})$ and the initial parameter vector is close to the true parameter vector.

\subsection{Derivation of the Rao test}\label{app:theory_Rao}
The Rao detector was originally proposed in \cite{rao1948large}, with a detailed derivation available in, for example, \cite{kay1998fundamentals}. 
Here, we present a slightly different derivation based on the local likelihood expansion introduced above. 
This serves to provide the reader with additional context, particularly in relation to the LRao detector discussed below.

To derive the Rao test statistic, we combine the local log-likelihood expansion with the likelihood ratio test, also known as the Neyman-Pearson test.
The log-likelihood ratio test statistic is given by
\begin{equation}
    \log T_{\mathrm{LRT}}= L(\boldsymbol{\theta}_1;\mathbf{x})-L_0.
    \label{equ:appLRT}
\end{equation}
Because $\boldsymbol{\theta}_1$ is unknown, the LMLE \eqref{equ:appLMLE} is used instead. Combining \eqref{equ:appTSELL}, \eqref{equ:appLMLE}, and \eqref{equ:appLRT} yields
\begin{align}
    \log T_{\mathrm{LRT}}&\approx L(\hat{\boldsymbol\theta}_{\mathrm{LMLE}};\mathbf{x})-L_0\\
    &\approx\nabla_{\boldsymbol\theta}L_0\mathbf{F}^{-1}_{0}(\nabla_{\boldsymbol\theta}L_0)^\top 
    -\frac{1}{2}\nabla_{\boldsymbol\theta}L_0\mathbf{F}^{-1}_{0}\mathbf{F}_0\mathbf{F}^{-1}_{0}(\nabla_{\boldsymbol\theta}L_0)^\top \\
    &=\frac{1}{2}\nabla_{\boldsymbol\theta}L_0\mathbf{F}^{-1}_{0}(\nabla_{\boldsymbol\theta}L_0)^\top .
\end{align}
The factor $1/2$ can be transferred into the decision threshold \cite{kay1998fundamentals} such that the Rao test statistic is given by
\begin{equation}
    T_{\mathrm{Rao}}=\nabla_{\boldsymbol\theta}L_0\mathbf{F}^{-1}_{0}(\nabla_{\boldsymbol\theta}L_0)^\top ,
\end{equation}
and its asymptotic distribution can be shown \cite[p. 239]{kay1998fundamentals} to be $T_{\mathrm{Rao}}\sim\chi_l^2$ under $\mathcal{H}_0$, and $T_{\mathrm{Rao}}\sim\chi_l'^{{2}}(\lambda_\mathrm{Rao})$ with $\lambda_\mathrm{Rao}=(\boldsymbol{\theta}_1-\boldsymbol{\theta}_0)^\top \mathbf{F}_0(\boldsymbol{\theta}_1-\boldsymbol{\theta}_0)$ under $\mathcal{H}_1$.
Hence, the FPR ($P_\mathrm{FA}$) and the TPR ($P_\mathrm{D}$) are asymptotically given by
\begin{align}
    \mathrm{FPR}&=Q_{\chi_l^2}(\gamma)\\
    \mathrm{TPR}&=Q_{\chi_l'^2(\lambda_\mathrm{Rao})}(\gamma),
\end{align}
where $Q$ denotes the right tail probability of the distribution indicated in the subscript.
The threshold $\gamma$ can be chosen to satisfy a given FPR constraint.
Since the PDF under $\mathcal{H}_0$ and therefore also the FPR are independent of the unknown parameter, the detector has the desirable constant false alarm rate (CFAR) property with respect to $\boldsymbol\theta$ \cite{van2013detection}.

It can easily be verified that the Rao test can equivalently be written as
\begin{equation}
    T_{\mathrm{Rao}}=(\hat{\boldsymbol\theta}_{\mathrm{LMLE}}-\boldsymbol\theta_0)^\top \mathbf{F}_{0}(\hat{\boldsymbol\theta}_{\mathrm{LMLE}}-\boldsymbol\theta_0).
\end{equation}
Intuitively, the Rao detector first estimates the unknown parameter vector with the LMLE (which is often much easier to evaluate than the MLE \cite{kay1998fundamentals}) and compares its squared deviation from the reference parameter with its expected variance, i.e., the inverse FI.
If the squared deviation is much larger than the expected variance of the estimator, the detection statistic will exceed a reasonably set threshold, and we decide for $\mathcal{H}_1$.
Conversely, if the squared deviation is similar to the expected variance, we conclude that the underlying parameter vector is indeed the reference parameter and we decide for $\mathcal{H}_0$.

\subsection{LFI and pessimistic PDF}\label{app:theory_pess}
The PDF $p(\mathbf{x};\boldsymbol\theta)$ is usually unknown, and one has to make assumptions about its form.
In \cite{stein2014lower}, the so-called pessimistic PDF occurs as a PDF whose FI provides a lower bound for the FI of the original PDF.
We refer to the FI of the pessimistic PDF as linear Fisher information (LFI) and denote it as $\mathbf{J}$.
Mathematically, the bound can be written as
\begin{equation}
    \mathbf{F}\geq\mathbf{J},
    \label{equ:app_FIbound}
\end{equation}
where the inequality means that $\mathbf{F}-\mathbf{J}$ is a positive semidefinite matrix.
Under the pessimistic PDF assumption, the data follows a Gaussian distribution with mean $\boldsymbol{\mu}(\boldsymbol\theta)$, which locally contains all the information about the parameter vector $\boldsymbol\theta$, and covariance matrix $\boldsymbol\Sigma$, which we assume to be independent of $\boldsymbol\theta$, locally.
The LFI is given by
\begin{equation}
    \mathbf{J}=(\nabla_{\boldsymbol\theta}\boldsymbol\mu)^\top \boldsymbol\Sigma^{-1}(\nabla_{\boldsymbol\theta}\boldsymbol\mu),
\end{equation}
and its inverse provides an attainable lower bound for the covariance matrix of a locally unbiased linear estimator $\hat{\boldsymbol{\theta}}_{\mathrm{lin}}$ (linear in the data vector $\mathbf{x}$) around some reference parameter vector $\boldsymbol{\theta}_0$ \cite{kohn2016correlations}, or
\begin{equation}
    \mathrm{Cov}[\hat{\boldsymbol{\theta}}_{\mathrm{lin}}]\geq \mathbf{J}_0^{-1}.
\end{equation}
Similar to the fact that invertible transformations do not change the FI \cite{zamir1998proof}, invertible affine transformations do not change the LFI, as will be shown in the following.

Let $\mathbf{y}=\mathbf{A}\mathbf{x}+\mathbf{b}$, where $\mathbf{A}$ is an invertible matrix.
Then $\boldsymbol{\mu}_\mathbf{y}=\mathbf{A}\boldsymbol{\mu}+\mathbf{b}$, and  $\boldsymbol{\Sigma}_\mathbf{y}=\mathbf{A}\boldsymbol{\Sigma}\mathbf{A}^\top $.
The LFI of $\mathbf{y}$ is given by
\begin{align}
    \mathbf{J}_\mathbf{y}&=(\nabla_{\boldsymbol{\theta}}(\mathbf{A}\boldsymbol{\mu}+\mathbf{b}))^\top (\mathbf{A}\boldsymbol{\Sigma}\mathbf{A}^\top )^{-1}(\nabla_{\boldsymbol{\theta}}(\mathbf{A}\boldsymbol{\mu}+\mathbf{b}))\\
    &=(\nabla_{\boldsymbol{\theta}}\boldsymbol{\mu})^\top \mathbf{A}^\top \mathbf{A}^{-\top}\boldsymbol{\Sigma}^{-1}\mathbf{A}^{-1}\mathbf{A}(\nabla_{\boldsymbol{\theta}}\boldsymbol{\mu})\\
    &=(\nabla_{\boldsymbol{\theta}}\boldsymbol{\mu})^\top \boldsymbol{\Sigma}^{-1}(\nabla_{\boldsymbol{\theta}}\boldsymbol{\mu})\\
    &=\mathbf{J}.
\end{align}
Thus, the LFI not only provides a lower bound for the FI, but its inverse also bounds the variance of a locally unbiased linear estimator, and similar to the FI that remains invariant under invertible transformations, the LFI remains invariant under invertible affine transformations.
In the following, we make use of a local Taylor series expansion to derive the LBLUE and the LRao detector, in a manner analogous to the derivations of the LMLE and the Rao detector.

We define the pessimistic log-likelihood function as $\tilde{L}=\log\tilde{p}(\mathbf{x};\boldsymbol\theta)$, with $\tilde{p}(\mathbf{x},\boldsymbol\theta)=\mathcal{N}(\boldsymbol{\mu}(\boldsymbol\theta),\boldsymbol\Sigma)$.
We assume that $\boldsymbol\Sigma$ is independent of $\boldsymbol\theta$.
This is not a restriction as we use the pessimistic PDF description only locally around some reference parameter vector $\boldsymbol\theta_0$ and evaluate $\boldsymbol\Sigma$ at $\boldsymbol\theta_0$, denoted as $\boldsymbol\Sigma_0$.
Proceeding as above, the second-order Taylor series expansion of $\tilde{L}$ around $\boldsymbol\theta_0$ is given by
\begin{equation}
    \tilde{L}\approx \tilde{L}_0+\nabla_{\boldsymbol\theta}\tilde{L}_0(\boldsymbol\theta-\boldsymbol\theta_0)
    +\frac{1}{2}(\boldsymbol\theta-\boldsymbol\theta_0)^\top \nabla_{\boldsymbol\theta}\nabla_{\boldsymbol\theta}^\top \tilde{L}_0(\boldsymbol\theta-\boldsymbol\theta_0).
    \label{equ:TSELL_pess}
\end{equation}
The pessimistic score function evaluated at $\boldsymbol\theta_0$ is given by
\begin{equation}
    \nabla_{\boldsymbol\theta}\tilde{L}_0=(\mathbf{x}-\boldsymbol\mu_0)^\top \boldsymbol\Sigma_0^{-1}\nabla_{\boldsymbol\theta}\boldsymbol\mu_0,
    \label{equ:lfiscore}
\end{equation}
and can be seen as a local linear sufficient statistic, similar to the score function, which is a local sufficient statistic \cite{alsing2018generalized}.
The pessimistic score function performs a down-projection from $N\to l$, while conserving the LFI, as can be deduced from the statistical properties of the LBLUE, as presented below.
The pessimistic Hessian is given by
\begin{align}
    \nabla_{\boldsymbol\theta}\nabla_{\boldsymbol\theta}^\top \tilde{L}_0&=-(\nabla_{\boldsymbol\theta}\boldsymbol\mu_0)^\top \boldsymbol\Sigma_0^{-1}(\nabla_{\boldsymbol\theta}\boldsymbol\mu_0)+(\mathbf{x}-\boldsymbol\mu_0)^\top \boldsymbol\Sigma_0^{-1}\nabla_{\boldsymbol\theta}\nabla_{\boldsymbol\theta}^\top \boldsymbol\mu_0\\
    &=-\mathbf{J}_0+(\mathbf{x}-\boldsymbol\mu_0)^\top \boldsymbol\Sigma_0^{-1}\nabla_{\boldsymbol\theta}\nabla_{\boldsymbol\theta}^\top \boldsymbol\mu_0.
\label{equ:hessian_pess_sys}
\end{align}
As above, replacing the Hessian in \eqref{equ:TSELL_pess} by its expectation, i.e., $\mathbb{E}[\nabla_{\boldsymbol\theta}^2\tilde{L}_0]=-\mathbf{J}_0$, yields
\begin{align}
    \tilde{L}&\approx \tilde{L}_0+(\mathbf{x}-\boldsymbol\mu_0)^\top \boldsymbol\Sigma_0^{-1}\nabla_{\boldsymbol\theta}\boldsymbol\mu_0(\boldsymbol\theta-\boldsymbol\theta_0)
    -\frac{1}{2}(\boldsymbol\theta-\boldsymbol\theta_0)^\top \mathbf{J}_0(\boldsymbol\theta-\boldsymbol\theta_0).
    \label{equ:app_TSELL_pess_LFI}
\end{align}

\subsection{Derivation of the LBLUE}\label{app:theory_LBLUE}
The LBLUE for $\boldsymbol{\theta}$ is the LMLE under the pessimistic PDF assumption, and can therefore be obtained from $\nabla_{\boldsymbol\theta}\Tilde{L}|_{\boldsymbol\theta_{\mathrm{LBLUE}}}=\mathbf{0}^\top $, where the local expansion of  $\nabla_{\boldsymbol\theta}\Tilde{L}$ is used.
Taking the derivative of \eqref{equ:app_TSELL_pess_LFI} with respect to $\boldsymbol\theta$ yields
\begin{equation}
    \nabla_{\boldsymbol\theta}\Tilde{L}\approx (\mathbf{x}-\boldsymbol\mu_0)^\top \boldsymbol\Sigma_0^{-1}\nabla_{\boldsymbol\theta}\boldsymbol\mu_0
    -(\boldsymbol\theta-\boldsymbol\theta_0)^\top \mathbf{J}_0.
\end{equation}
Setting to zero and solving for $\boldsymbol{\theta}$ leads to
\begin{align}
    \hat{\boldsymbol\theta}_{\mathrm{LBLUE}}=\boldsymbol\theta_0+\mathbf{J}^{-1}_{0}(\nabla_{\boldsymbol\theta}\boldsymbol\mu_0)^\top \boldsymbol\Sigma_0^{-1}(\mathbf{x}-\boldsymbol\mu_0).
    \label{equ:app_LBLUE}
\end{align}

We are now going to verify our assumption that the statistic of the LBLUE is asymptotically Gaussian, with a mean corresponding to the parameter $\boldsymbol{\theta}$ and a covariance matrix equal to the inverse LFI $\mathbf{J}_0^{-1}$, i.e., $\hat{\boldsymbol\theta}_{\mathrm{LBLUE}}\overset{a}{\sim}\mathcal{N}(\boldsymbol\theta,\mathbf{J}_0^{-1})$, where $a$ indicates that this relation holds asymptotically for an infinitely large sample size $N\to\infty$.

We start by deriving the mean and covariance of the LBLUE.
Using $\boldsymbol\mu\approx \boldsymbol\mu_0+\nabla_{\boldsymbol{\theta}}\boldsymbol{\mu}_0(\boldsymbol{\theta}-\boldsymbol{\theta}_0)$, we find
\begin{equation}
\begin{split}
    \mathbb{E}[\hat{\boldsymbol\theta}_{\mathrm{LBLUE}}]&\approx\boldsymbol\theta_0+\mathbf{J}^{-1}_0(\nabla_{\boldsymbol{\theta}}\boldsymbol{\mu}_0)^\top \boldsymbol{\Sigma}_0^{-1}(\boldsymbol\mu_0+\nabla_{\boldsymbol{\theta}}\boldsymbol{\mu}_0(\boldsymbol{\theta}-\boldsymbol{\theta}_0)-\boldsymbol\mu_0)\\
    &=\boldsymbol\theta_0+\mathbf{J}^{-1}_0\mathbf{J}_0(\boldsymbol{\theta}-\boldsymbol{\theta}_0)\\
    &=\boldsymbol\theta,
    \label{equ:app_lblue_exp}
\end{split}
\end{equation}
and
\begin{equation}
\begin{split}
    \mathrm{Cov}[\hat{\boldsymbol\theta}_{\mathrm{LBLUE}}]&=\mathbf{J}^{-1}_0(\nabla_{\boldsymbol{\theta}}\boldsymbol{\mu}_0)^\top \boldsymbol{\Sigma}_0^{-1}\mathrm{Cov}[\mathbf{x}-\boldsymbol\mu_0]\boldsymbol{\Sigma}_0^{-1}(\nabla_{\boldsymbol{\theta}}\boldsymbol{\mu}_0)\mathbf{J}^{-1}_0\\
    &=\mathbf{J}^{-1}_0(\nabla_{\boldsymbol{\theta}}\boldsymbol{\mu}_0)^\top \boldsymbol{\Sigma}_0^{-1}(\nabla_{\boldsymbol{\theta}}\boldsymbol{\mu}_0)\mathbf{J}^{-1}_0\\
    &=\mathbf{J}^{-1}_0\mathbf{J}_0\mathbf{J}^{-1}_0\\
    &=\mathbf{J}^{-1}_0.
    \label{equ:app_lblue_var}
\end{split}
\end{equation}
The remaining task is to verify that the distribution of $\hat{\boldsymbol\theta}_{\mathrm{LBLUE}}$ is indeed asymptotically Gaussian. 
Since the entries of the random variable $\mathbf{x}$ are not independent, asymptotic normality does not immediately follow from the conventional central limit theorem, which requires independence. 
However, when dealing with time series data, the values of a random variable typically only depend on each other if they are sufficiently close in time.
Here, we assume $m$-dependence \cite{orey1958central}, which means that components $x_i$ and $x_j$ of the random variable $\mathbf{x}$ are dependent if $|i-j|\leq m$ and independent if $|i-j|> m$. \cite{orey1958central} provides a proof of a suitable version of the central limit theorem for $m$-dependent random variables that immediately implies asymptotic normality of $\hat{\boldsymbol\theta}_{\mathrm{LBLUE}}$.

Assuming $m$-dependence is not the only way to obtain normality of the LBLUE.
In general, there exists a plethora of conditions on the dependence between the random variables' components, and as a consequence, multiple variants of the central limit theorem for these weakly dependent random variables. 
In any case, one should analyze the dependencies between the components when applying our method of signal detection. 
While the exact nature of the dependencies is difficult to infer just from the data, we can get a good intuition about how the entries of the random vector depend on each other by (robust) estimation of the covariance matrix and comparing its diagonal elements with the elements that are far off the diagonal. 
For stationary data, one can estimate the autocorrelation sequence robustly to visually analyze the dependence.
A more advanced and computationally demanding way of directly assessing the dependencies between the components would be to estimate the mutual information between them \cite{kraskov2004estimating}.

Adding to the discussion on dependence, we expect that the statistic of the LBLUE can be modeled to a good approximation by a Gaussian distribution. 
Hence, as a practical alternative to carefully measuring the dependencies, one can estimate the PDF of the LBLUE in order to confirm that it is Gaussian.

\subsection{Derivation of the LRao test}\label{app:theory_LRao}
In the following, we present a derivation of the proposed LRao test.
As noted in the main text, the LRao test can be obtained simply by evaluating the Rao test under the pessimistic PDF assumption. 
Here, we take a slightly deeper approach, starting with the log-likelihood ratio test statistic under this pessimistic PDF assumption
\begin{equation}
    \log \widetilde{T}_{\mathrm{LRT}}= \Tilde{L}(\boldsymbol{\theta}_1;\mathbf{x})-\Tilde{L}_0.
    \label{equ:app_pess_LRT}
\end{equation}
Similar to above, we insert the LBLUE \eqref{equ:app_LBLUE} for $\boldsymbol{\theta}_1$ as it is unknown.
Combining \eqref{equ:app_TSELL_pess_LFI}, \eqref{equ:app_LBLUE}, and \eqref{equ:app_pess_LRT} yields
\begin{align}
    \log \widetilde{T}_{\mathrm{LRT}}&\approx \Tilde{L}(\hat{\boldsymbol\theta}_{\mathrm{LBLUE}})-\Tilde{L}(\boldsymbol\theta_0)\\
    &=\frac{1}{2}(\mathbf{x}-\boldsymbol\mu_0)^\top \boldsymbol\Sigma_0^{-1}(\nabla_{\boldsymbol\theta}\boldsymbol\mu_0)\mathbf{J}^{-1}_{0}(\nabla_{\boldsymbol\theta}\boldsymbol\mu_0)^\top \boldsymbol\Sigma_0^{-1}(\mathbf{x}-\boldsymbol\mu_0).
\end{align}
As above, the factor can be transferred to the threshold such that the LRao statistic is given by
\begin{align}
    T_{\mathrm{LRao}}&=\nabla_{\boldsymbol\theta}\Tilde{L}_0\mathbf{J}^{-1}_{0}(\nabla_{\boldsymbol\theta}\Tilde{L}_0)^\top \\
    &=(\mathbf{x}-\boldsymbol\mu_0)^\top \boldsymbol\Sigma_0^{-1}(\nabla_{\boldsymbol\theta}\boldsymbol\mu_0)\mathbf{J}^{-1}_{0}(\nabla_{\boldsymbol\theta}\boldsymbol\mu_0)^\top \boldsymbol\Sigma_0^{-1}(\mathbf{x}-\boldsymbol\mu_0).
\end{align}
Its asymptotic distribution can be derived as follows.
It is easily verified that the LRao test statistic can be rewritten as
\begin{equation}
    T_{\mathrm{LRao}}=(\hat{\boldsymbol\theta}_{\mathrm{LBLUE}}-\boldsymbol\theta_0)^\top \mathbf{J}_{0}(\hat{\boldsymbol\theta}_{\mathrm{LBLUE}}-\boldsymbol\theta_0).
\end{equation}
Assuming that the LBLUE attains its asymptotic distribution $\mathcal{N}(\boldsymbol\theta,\mathbf{J}_0^{-1})$, it directly follows \cite[p.32, f.]{kay1998fundamentals} that $T_{\mathrm{LRao}}\sim\chi_l^2$ under $\mathcal{H}_0$, and $T_{\mathrm{LRao}}\sim\chi_l'^{{2}}(\lambda_\mathrm{LRao})$ with $\lambda_\mathrm{LRao}=(\boldsymbol{\theta}_1-\boldsymbol{\theta})^\top \mathbf{J}_0(\boldsymbol{\theta}_1-\boldsymbol{\theta})$ under $\mathcal{H}_1$.
Hence, the FPR ($P_\mathrm{FA}$) and the TPR ($P_\mathrm{D}$) are given by
\begin{align}
    \mathrm{FPR}&=Q_{\chi_l^2}(\gamma)\\
    \mathrm{TPR}&=Q_{\chi_l'^2(\lambda_\mathrm{LRao})}(\gamma),
\end{align}
where $Q$ denotes the right tail probability of the distribution indicated in the subscript.
Similar to the Rao test, the threshold $\gamma$ can be chosen to satisfy a given FPR constraint.
Since the PDF under $\mathcal{H}_0$ and therefore also the FPR are independent of the unknown parameter, the detector has the desirable constant false alarm rate (CFAR) property with respect to $\boldsymbol\theta$.

\subsection{Proof that the score function maximizes the trace of LFI}\label{app:theory_LFImaxscore}
While it was recently noted in \cite{weimar2025fisher} that score-transformed data maximizes the LFI, we show here that for vector parameters, it maximizes the trace of the LFI and present a formal proof.
The LFI of the data transformed by the score function $\mathbf{x}'=(\nabla_{\boldsymbol{\theta}}L_0)^\top$ (which does not depend on $\boldsymbol\theta$) is defined as
\begin{equation}
    \mathbf{J}_0'=(\nabla_{\boldsymbol{\theta}}\boldsymbol\mu_0')^\top{\boldsymbol\Sigma_0'}^{-1}(\nabla_{\boldsymbol{\theta}}\boldsymbol\mu_0').
    \label{equ:app_lfi_score}
\end{equation}
Starting with the gradient of the mean, we get
\begin{align}
    \nabla_{\boldsymbol{\theta}}\boldsymbol\mu_0'&=(\nabla_{\boldsymbol{\theta}}\mathbb{E}[(\nabla_{\boldsymbol{\theta}}L_0)^\top])|_{\boldsymbol\theta=\boldsymbol\theta_0}\\
    &=\int(\nabla_{\boldsymbol{\theta}}L_0)^\top\nabla_{\boldsymbol{\theta}} p(\mathbf{x};\boldsymbol\theta)\mathrm{d}\mathbf{x}|_{\boldsymbol\theta=\boldsymbol\theta_0}\\
    &=\int(\nabla_{\boldsymbol{\theta}}L_0)^\top\frac{\nabla_{\boldsymbol{\theta}} p(\mathbf{x};\boldsymbol\theta)}{p(\mathbf{x};\boldsymbol\theta)}p(\mathbf{x};\boldsymbol\theta)\mathrm{d}\mathbf{x}|_{\boldsymbol\theta=\boldsymbol\theta_0}\\
    &=\int(\nabla_{\boldsymbol{\theta}}L_0)^\top(\nabla_{\boldsymbol{\theta}} L)p(\mathbf{x};\boldsymbol\theta)\mathrm{d}\mathbf{x}|_{\boldsymbol\theta=\boldsymbol\theta_0}\\
    &=\int(\nabla_{\boldsymbol{\theta}}L_0)^\top(\nabla_{\boldsymbol{\theta}} L_0)p(\mathbf{x};\boldsymbol\theta_0)\mathrm{d}\mathbf{x}\\
    &=\mathbf{F}_0.
    \label{equ:app_lfi_score_mean}
\end{align}
Proceeding with the covariance matrix, we obtain
\begin{align}
    {\boldsymbol\Sigma_0'}&=\mathbb{E}[(\nabla_{\boldsymbol{\theta}}L_0-\mathbb{E}[\nabla_{\boldsymbol{\theta}}L_0])^\top(\nabla_{\boldsymbol{\theta}}L_0-\mathbb{E}[\nabla_{\boldsymbol{\theta}}L_0])]_{\boldsymbol\theta=\boldsymbol\theta_0}\\
    &=\mathbb{E}[(\nabla_{\boldsymbol{\theta}}L_0-\mathbb{E}[\nabla_{\boldsymbol{\theta}}L_0]_{\boldsymbol\theta=\boldsymbol\theta_0})^\top(\nabla_{\boldsymbol{\theta}}L_0-\mathbb{E}[\nabla_{\boldsymbol{\theta}}L_0]_{\boldsymbol\theta=\boldsymbol\theta_0})]_{\boldsymbol\theta=\boldsymbol\theta_0}\\
    &=\mathbb{E}[(\nabla_{\boldsymbol{\theta}}L_0)^\top(\nabla_{\boldsymbol{\theta}}L_0)]_{\boldsymbol\theta=\boldsymbol\theta_0}\\
    &=\mathbf{F}_0.
    \label{equ:app_lfi_score_cov}
\end{align}
where we used the regularity condition $\mathbb{E}[(\nabla_{\boldsymbol{\theta}}L_0)^\top]_{\boldsymbol\theta=\boldsymbol\theta_0}=\mathbf{0}$ \cite{kay1993fundamentals}.
Inserting \eqref{equ:app_lfi_score_mean} and \eqref{equ:app_lfi_score_cov} into \eqref{equ:app_lfi_score} yields
\begin{equation}
    \mathbf{J}_0'=\mathbf{F}_0\mathbf{F}_0^{-1}\mathbf{F}_0=\mathbf{F}_0.
    \label{equ:app_lfi_score_inserted}
\end{equation}
Due to \eqref{equ:app_FIbound}, this proves that the LFI and its trace are maximized by the score function.
Moreover, every affine transformation of the score function maximizes the LFI.
This follows from the fact that invertible affine transformations do not change the LFI, as was shown in SI~\ref{app:theory_pess}.

Finally, we note that, similar to the maximization of the trace of the LFI, score matching, which has recently been very successful in the field of generative AI, also produces the score function at its optimum \cite{hyvarinen2005estimation,bishop2023deep}.
As generative models learn the distribution of the data, score matching produces usually focuses on the so-called Stein score, defined as $\nabla_\mathrm{x} L$ rather than the Fisher score $\nabla_{\boldsymbol{\theta}} L$.

\subsection{One sided local parameter testing}\label{app:theory_LLMP}
Our method of detection in the LFI framework with the LRao test can also be adapted to one-sided parameter testing.
While the resulting statistic (differently scaled and without the LFI optimization) was first derived in \cite{moreno2014information}, we present it here, to the best of our knowledge, for the first time in the context of detection theory.

We consider the usual scalar one-sided test, where the hypotheses are defined as
\begin{eqnarray}
    &\mathcal{H}_0:\ \theta=\theta_0&\\
    &\mathcal{H}_1:\ \theta>\theta_0&.
\end{eqnarray}
The locally most powerful (LMP) test statistic, which is locally optimal \cite[p.217]{kay1998fundamentals}, is given by 
\begin{equation}
    T_\mathrm{LMP}=\frac{1}{\sqrt{F_0}}\frac{\partial L}{\partial\theta}\bigg|_{\theta=\theta_0},
\end{equation}
and its asymptotic distribution is $\mathcal{N}(0,1)$ under $\mathcal{H}_0$, and $\mathcal{N}(\sqrt{F_0}(\theta_1-\theta_0),1)$ under $\mathcal{H}_1$ \cite{kay1998fundamentals}.
This detector can be interpreted as a one-sided Rao test, where the squaring operation is omitted, since $\theta\geq\theta_0$.
Extending the LMP test to the LFI framework yields the linear locally most powerful (LLMP) test statistic (one could also say one-sided LRao test, but we call it LLMP test to be consistent with the literature's LMP test \cite{kay1998fundamentals})
\begin{equation}
    T_\mathrm{LLMP}=\frac{1}{\sqrt{J_0}}\frac{\partial\boldsymbol{\mu}^\top}{\partial\theta}\bigg|_{\theta=\theta_0}\boldsymbol{\Sigma}_0^{-1}(\mathbf{x}-\boldsymbol\mu_0),
\end{equation}
and its asymptotic distribution is $\mathcal{N}(0,1)$ under $\mathcal{H}_0$, and $\mathcal{N}(\sqrt{J_0}(\theta_1-\theta_0),1)$ under $\mathcal{H}_1$, which directly follows from the asymptotic LBLUE statistic.


\section{Weak signal detection}\label{app:weak}
This part of the SI~covers the details of our method related to Section \ref{sec:weak}.

\subsection{Additive signal model}\label{app:weak_additive}
Within the scope of this work, we assume that our data can be accurately modeled by an additive noise model
\begin{equation}
    \mathbf{x} = \mathbf{f}(\boldsymbol{\theta}) + \mathbf{w}\;,
    \label{app:additive_noise}
\end{equation}
where $\mathbf{f}(\boldsymbol{\theta})$ is a known deterministic function and $\mathbf{w}\sim p_\mathbf{w}(\mathbf{w})$ is a stochastic term. While in this model we assume that the noise is independent of the deterministic contribution, it allows for an arbitrarily complex distribution of the noise. Note that the assumption of additive noise is not a strict requirement, and many of our results do still apply to other models or can be generalized accordingly. Assuming additive noise is particularly useful for the task of weak signal detection, as we show in the following.

Detecting weak signal means that we want to differentiate if the data comes from a distribution with the parameter $\boldsymbol{\theta}_0$ or with the parameter $\boldsymbol{\theta}_0 + \delta \boldsymbol{\theta}$, where the magnitudes of the components of $\delta \boldsymbol{\theta}$ are small. This implies that we can linearize the function
\begin{equation}
    \mathbf{f}(\boldsymbol{\theta})\approx \mathbf{f}(\boldsymbol{\theta}_0) +  \nabla_{\boldsymbol\theta}\mathbf{f}(\boldsymbol{\theta}_0)(\boldsymbol{\theta}-\boldsymbol{\theta}_0)\;,
\end{equation}
where $\nabla_{\boldsymbol\theta}\mathbf{f}(\boldsymbol{\theta}_0)$ is the Jacobi matrix of $\mathbf{f}$ evaluated at $\boldsymbol\theta_0$.
Without loss of generality, we can subtract the offset $\mathbf{f}(\boldsymbol\theta_0)-\nabla_{\boldsymbol\theta_0}\mathbf{f}(\boldsymbol\theta)\boldsymbol\theta_0$ from the data and substitute $\boldsymbol\theta$ for $\boldsymbol\theta-\boldsymbol\theta_0$ to arrive at \eqref{equ:additive_signal_model}, or $\mathbf{f}(\boldsymbol\theta)\approx\mathbf{H}\boldsymbol\theta$ with $\boldsymbol\theta_0=\mathbf{0}$.

Moreover, additive noise enables us to calculate the derivatives with respect to the parameter in post-processing, without actually shifting the parameter by small amounts in the experiment, which provides us with a significant practical advantage.
In order to evaluate the LFI, the LBLUE, and the LRao test, we are interested in the Jacobian $\nabla_{\boldsymbol\theta}\boldsymbol\mu_{\Psi,0}$ at the output of a neural network $\Psi_W(\mathbf{x})$.
Specifically, the $j$th column of the Jacobian can be approximated by the central difference quotient
\begin{equation}
    \frac{\partial\boldsymbol\mu_{\Psi,0}}{\partial\theta_j}\approx\sum_i\frac{\Psi_W(\mathbf{w}_i+\mathbf{h}_j\Delta\theta)-\Psi_W(\mathbf{w}_i-\mathbf{h}_j\Delta\theta)}{2\Delta\theta}
    \label{equ:app_dq}
\end{equation}
where $\mathbf{w}_i$ is the $i$th training data (noise-only) sequence, $\mathbf{h}_j$ is the $j$th column of $\mathbf{H}$, and $\Delta\theta$ is a step size parameter, which is discussed thoroughly in SI~\ref{app:weak_neural}.

An additional advantage of the additive model \eqref{equ:additive_signal_model} is that it allows for a convenient generalization to the case where $\mathbf{H}$ is unknown a priori, or contains additional unknown parameters, i.e., nuisance parameters which are not present under $\mathcal{H}_0$.
We can optimize our neural network under the assumption of a different data model $\mathbf{x}=\boldsymbol\theta^*+\mathbf{w}$, and do the multiplication by $\mathbf{H}$ afterwards.
This result can be obtained by applying the chain rule as follows.
The Jacobian of the mean at the output of a neural network is given by
\begin{equation}
    \begin{split}
    \nabla_{\boldsymbol\theta}\boldsymbol\mu_{\Psi,0}&=\int\Psi_W(\mathbf{x})\nabla_{\boldsymbol\theta}p_\mathbf{w}(\mathbf{x}-\mathbf{H}\boldsymbol\theta)|_{\boldsymbol\theta=\mathbf{0}}\mathrm{d}\mathbf{x}\\
    &=\int\Psi_W(\mathbf{x})\nabla_{\boldsymbol\theta^*} p_\mathbf{w}(\mathbf{x}-\boldsymbol\theta^*)|_{\boldsymbol\theta^*=\mathbf{0}}\mathrm{d}\mathbf{x}\mathbf{H}\\
    &=\nabla_{\boldsymbol\theta^*}\boldsymbol\mu_{\Psi,0}\mathbf{H},
    \end{split}
\end{equation}
and the LFI is given by
\begin{equation}
    \begin{split}
    \mathbf{J}_{\Psi,0}&=(\nabla_{\boldsymbol\theta}\boldsymbol\mu_{\Psi,0})^\top\boldsymbol\Sigma_{\Psi,0}^{-1}(\nabla_{\boldsymbol\theta}\boldsymbol\mu_{\Psi,0})\\
    &=\mathbf{H}^\top(\nabla_{\boldsymbol\theta^*}\boldsymbol\mu_{\Psi,0})^\top\boldsymbol\Sigma_{\Psi,0}^{-1}(\nabla_{\boldsymbol\theta^*}\boldsymbol\mu_{\Psi,0})\mathbf{H}\\
    &=\mathbf{H}^\top\mathbf{J}_{\Psi,0}^*\mathbf{H},
    \end{split}
\end{equation}
where $\mathbf{J}_{\Psi,0}^*=(\nabla_{\boldsymbol\theta^*}\boldsymbol\mu_{\Psi,0})^\top\boldsymbol\Sigma_{\Psi,0}^{-1}(\nabla_{\boldsymbol\theta^*}\boldsymbol\mu_{\Psi,0})$ is the LFI at the output of the neural network associated with the data model $\mathbf{x}=\boldsymbol\theta^*+\mathbf{w}$.
As above, the expression $\nabla_{\boldsymbol\theta^*}\boldsymbol\mu_{\Psi,0}$ can be computed in post-processing without physically shifting any parameter value.
Thus, if $\mathbf{H}$ is unknown a priori, or contains additional unknown parameters, we can optimize the network by maximizing the trace of $\mathbf{J}_{\Psi}^*$ using noise only data, and afterwards apply it to an arbitrary observation matrix $\mathbf{H}$.
A related quantity to $\mathbf{J}_{\Psi,0}^*$ for the scalar parameter case is the so-called intrinsic accuracy of a noise distribution \cite{kay1998fundamentals}.
It measures the amount of FI with respect to an additive shift parameter and plays a central role in weak signal detection under additive IID noise.


\subsection{Stationary data}\label{app:weak_stationary}
Stationarity is a central assumption in many areas of signal processing. 
It is often fulfilled in practice to a good approximation, which allows access to the highly useful tools \cite{papoulis2002probability}.
Among them are frequency domain methods, which come with the advantage that they often can be implemented using a computationally efficient fast Fourier transform (FFT) algorithm \cite{cooley1965algorithm}.
In the following, we describe how we arrive at relatively simple expressions for the LFI and its trace under the stationarity assumption.

The covariance matrix $\boldsymbol\Sigma$ of a random vector $\mathbf{x}$ drawn from a stationary process $x_n$ \cite{kay2006intuitive} has Toeplitz structure \cite{gray2006toeplitz}, because the correlation between two samples depends only on their distance.
Assuming that $x_n$ and $x_l$ are uncorrelated for $|n-l|>m$, where $N\gg m$ ($N$ is the dimension of $\boldsymbol\Sigma$), the eigenvalues of $\boldsymbol\Sigma$ can be approximated as
\begin{equation}
    \lambda_i\approx P_{xx}(f_i),
\end{equation}
where
\begin{equation}
    P_{xx}(f_i)=\sum\limits_{k=-\infty}^{\infty}r_k\mathrm{e}^{-\mathrm{j}2\pi f_ik}
\end{equation}
is the power spectral density sampled at the normalized DFT frequencies $f_i=i/N$ for $i=0,1,...,N-1$, $r_k=\mathbb{E}[x_nx_{n-k}]$ is the auto-covariance sequence (assuming the random process is zero mean), and its eigenvectors
\begin{align}
    \mathbf{v}_i=\frac{1}{\sqrt{N}}\begin{bmatrix}
        1 & \mathrm{e}^{\mathrm{j}2\pi f_i} & \mathrm{e}^{\mathrm{j}4\pi f_i} & \hdots & \mathrm{e}^{\mathrm{j}2\pi (N-1)f_i}
    \end{bmatrix}^\top 
\end{align}
align with the DFT vectors \cite{gray1972asymptotic}.
An intuitive explanation for this eigendecomposition can be found in \cite[p. 33, ff.]{kay1998fundamentals}
The covariance matrix can thus be approximated as
\begin{equation}
    \boldsymbol\Sigma\approx\sum\limits_{n=0}^{N-1}P_{xx}(f_i)\mathbf{v}_i\mathbf{v}_i^H,
\end{equation}
and its inverse as
\begin{equation}
    \boldsymbol\Sigma^{-1}\approx\sum\limits_{n=0}^{N-1}\frac{\mathbf{v}_i\mathbf{v}_i^H}{P_{xx}(f_i)}.
    \label{equ:app_asympt_cov_inv}
\end{equation}
Estimating the covariance matrix from data then boils down to power spectral density estimation since the eigenvectors are known.
This approximation comes with the significant advantage that the power spectral density only has $N/2$ unknown parameters (real data vectors have symmetric power spectral densities) as opposed to the $N(N-1)/2$ unknown parameters of the sample covariance matrix.
A vast amount of literature on power spectral density estimation exists \cite{stoica2005spectral,kay1988modern,proakis2007digital}.
We use one of the most prominent non-parametric techniques, the so-called Schuster periodogram \cite{schuster1898investigation}
\begin{equation}
    \hat{P}_{xx}(f_i)=\left|\sum\limits_{n=0}^{N-1}x_n\mathrm{e}^{-\mathrm{j}2\pi f_in}\right|^2.
\end{equation}
To reduce the variance of the estimate, the periodograms of multiple data vectors are averaged, similar in spirit to Bartlett's estimator \cite{bartlett1948smoothing}.

Using \eqref{equ:app_asympt_cov_inv}, the LFI can be expressed as
\begin{align}
    \mathbf{J}&=(\nabla_{\boldsymbol{\theta}}\boldsymbol{\mu})^\top \boldsymbol{\Sigma}^{-1}(\nabla_{\boldsymbol{\theta}}\boldsymbol{\mu})\\
    &\approx\sum\limits_{i=0}^{N-1}\frac{(\nabla_{\boldsymbol{\theta}}\boldsymbol{\mu})^\top \mathbf{v}_i\mathbf{v}_i^H\nabla_{\boldsymbol{\theta}}\boldsymbol{\mu}}{P_{xx}(f_i)},
\end{align}
where $\mathbf{v}_i^H\nabla_{\boldsymbol{\theta}}\boldsymbol\mu$ is the $i$th entry of the DFT of the Jacobian over the temporal dimension.
Calculating the trace of this approximation yields
\begin{equation}
     \mathrm{tr}(\mathbf{J})\approx\sum\limits_{j=1}^{l}\sum\limits_{i=0}^{N}\frac{|\mathbf{v}_i^H\frac{\partial\boldsymbol{\mu}}{\partial\theta_j}|^2}{{P_{xx}(f_i)}}.
    \label{equ:ap_cost_function}
\end{equation}
Both $P_{xx}(f_i)$ and $\mathbf{v}_i^H\nabla_{\boldsymbol{\theta}}\boldsymbol\mu$ can be estimated using an FFT algorithm.

\subsection{Periodic signal model}\label{app:weak_periodic}
For the practical example of weak periodic signal detection in magnetic sensor data, we use the additive signal model \eqref{equ:additive_signal_model}.
Periodic signals are commonly modeled using the Fourier series \cite{proakis2007digital}.
For finite $N$, the multi-harmonic signal model is typically used \cite{christensen2009multi,kay2013fundamentals} and is given by
\begin{equation}
    s_n=\sum\limits_{k=1}^{K}A_k\cos(2\pi kf_0n+\phi_k),
\end{equation}
for $n=0,1,\dots,N-1$, where $N=128$ is the signal length, $A_k>0$ and $-\pi\leq\phi_k<\pi$ are the amplitudes and phases, $f_0=0.1$ is the normalized fundamental frequency, and $K=4$ is the model order.
The signal model can be brought into the form of \eqref{equ:additive_signal_model} by writing
\begin{equation}
    s_n=\sum\limits_{k=1}^{K}\alpha_k\cos(2\pi kf_0n)+\beta_k\sin(2\pi kf_0n),
\end{equation}
where $\alpha_k=A_k\cos(\phi_k)$ and $\beta_k=-A_k\sin(\phi_k)$, or in matrix vector notation
\begin{equation}
    \mathbf{s}=\mathbf{H}\boldsymbol{\theta},
\end{equation}
with
\begin{equation}
    \mathbf{H}=\begin{bmatrix}
        \cos(2\pi f_00) & \hdots & \cos(2\pi f_0(N-1))\\
        \vdots & & \vdots\\
        \cos(2\pi Kf_00) & \hdots & \cos(2\pi Kf_0(N-1))\\
        \sin(2\pi f_00) & \hdots & \sin(2\pi f_0(N-1))\\
        \vdots & & \vdots\\
        \sin(2\pi Kf_00) & \hdots & \sin(2\pi Kf_0(N-1))\\
    \end{bmatrix}^\top 
\end{equation}
and
\begin{equation}
    \boldsymbol{\theta}=\begin{bmatrix}
        \alpha_1 & \hdots & \alpha_K & \beta_1 & \hdots & \beta_K
    \end{bmatrix}^\top .
\end{equation}

\subsection{Cauchy noise}\label{app:cauchy}
The Cauchy noise sequences for the simulation study are generated by generating IID vectors of Cauchy noise $\mathbf{z}\in\mathbb{R}^{N}$, where $z_i\sim\mathcal{C}(0,1)$ for $0\leq i<N$.
We then apply a linear filter to introduce dependency between closely spaced samples.
In particular, we use a randomly generated stable autoregressive filter of order 3.
The filtered noise sequence is given by $\mathbf{w}=\mathbf{A}^{-1}\mathbf{z}$, where $\mathbf{A}$ is a lower triangular filter matrix.
To derive the FI for the additive signal model \eqref{equ:additive_signal_model}, we multiply the equation by $\mathbf{A}$ from the left side to obtain $\mathbf{A}\mathbf{x}=\mathbf{A}\mathbf{H}\boldsymbol\theta+\mathbf{z}$, which does not alter the FI since $\mathbf{A}$ is invertible.
The FI with respect to $\boldsymbol\theta$ is then given by
\begin{equation}
    \mathbf{F}=\mathbf{H}^\top\mathbf{A}^\top\mathbf{F}^{*}\mathbf{A}\mathbf{H},
\end{equation}
where $\mathbf{F}^{*}=\frac{1}{2}\mathbf{I}$ is the FI corresponding to the model $\mathbf{x}^{*}=\boldsymbol\theta^*+\mathbf{z}$.

\subsection{Magnetic sensor data}\label{app:weak_magnetic}
In Fig.~\ref{fig:data_set}, the magnetic sensor data taken from the supplementary material in \cite{kay2013fundamentals} is shown.
The data contains a noise-only time series of length \num{10000} samples.
As a pre-processing step, we apply the robust shifting and normalization transformation
\begin{equation}
    \frac{\mathbf{x}-\mathrm{median}(\mathbf{x})}{1.483\cdot\mathrm{median}(|\mathbf{x}-\mathrm{median}(\mathbf{x})|)},
    \label{app:prenormshift}
\end{equation}
where the factor \num{1.483} is introduced to ensure consistency in the scale estimate for the standard deviation in a normal distribution \cite{zoubir2012robust,zoubir2018robust}.
We then split the data into non-overlapping sequences of length $N=128$ to obtain a total of $78$ data vectors used for training, validation, and testing.
Due to the limited amount of data available, a nested cross-validation procedure is applied, where the neural network weights are optimized on the training data, the hyperparameters are optimized on the validation data to prevent information leakage to the model \cite{cawley2010over}, and the performance is evaluated on the test data.
The outer cross-validation procedure (test $|$ train + validate) is 10-fold, the inner (train $|$ validate) is 9-fold.
When an inner cross-validation run is finished, the training and validation sets are combined to train the network with optimized hyperparameters and evaluate it on the test set.
The training and evaluation are repeated 100 times with randomly initialized weights (only in case of our network-based method, as the reference methods do not require initialization) and randomly selected generated $\mathcal{H}_1$ data as described next.

The $\mathcal{H}_1$ data for testing is obtained by adding generated periodic signals, see SI~\ref{app:weak_periodic}, to the test data sequences.
For simplicity, all amplitudes are chosen to be equal in value, depending on the signal-to-noise ratio used for the simulations.
The phases are drawn randomly from a uniform distribution $\mathcal{U}(-\pi,\pi)$.
This way, the signal is purely model-based, while the noise comes from real-world measurements.

\begin{figure}
    \centering
    \includegraphics[width=0.5\linewidth]{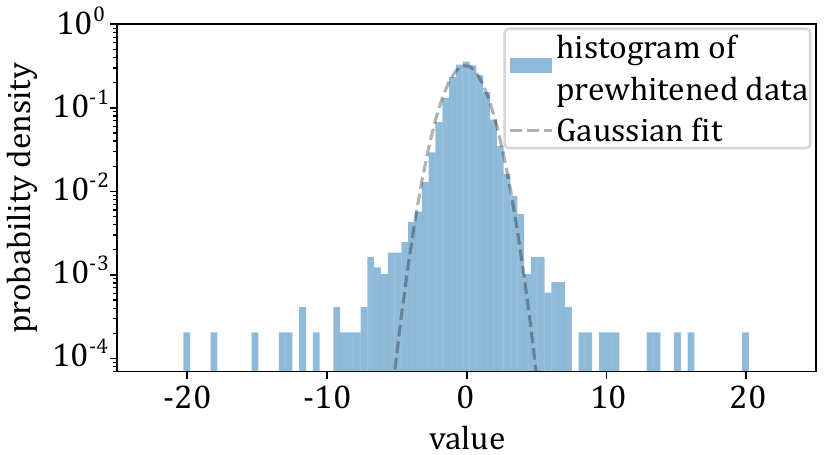}
    \caption{Normalized histogram and Gaussian fit of the prewhitened magnetic sensor data. Parts of the data can be well approximated by a Gaussian distribution.}
    \label{fig:hist_magn_data_Gauss}
\end{figure}

We now analyze the magnetic sensor data, which are plotted in Fig.~\ref{fig:data_set}, to motivate our choice of the reference methods and neural network hyperparameters.
Starting with the marginal data distribution, in Fig.~\ref{fig:hist_magn_data_Gauss}, the histogram of the prewhitened data is compared to a Gaussian fit.
The prewhitening procedure is described in SI~\ref{app:weak_ref}.
When neglecting the outliers, most of the data is well approximated by a Gaussian distribution.
However, for algorithms derived under the Gaussian noise assumption, even one single outlier can lead to severe performance degradation \cite{zoubir2012robust}.
For this reason, the reference methods with a nonlinear limiting function perform significantly better than the method without nonlinearity, as can be seen in Fig.~\ref{fig:data_set}.

\begin{figure}
    \centering
    \includegraphics[width=0.7\linewidth]{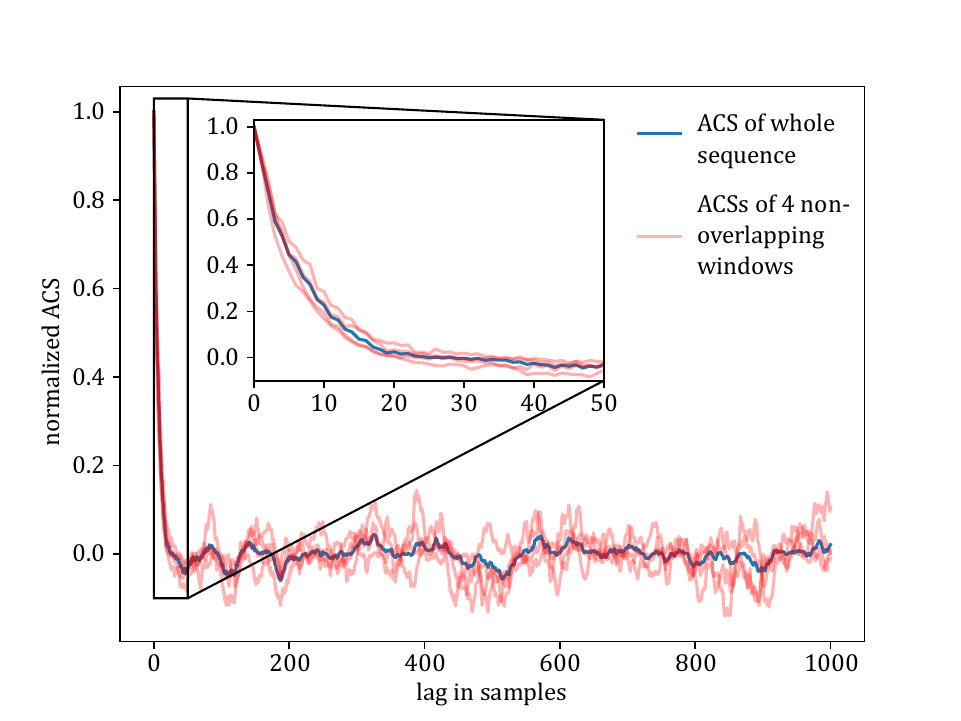}
    \caption{The robustly estimated auto-correlation sequence (ACS) of the magnetic sensor data. In blue, the ACS of the whole data sequence, and in red, the ACSs of 4 data windows that are obtained from splitting the original sequence into 4 non-overlapping consecutive parts are plotted. The data appear to be second-order weakly dependent and stationary.}
    \label{fig:app_ACS}
\end{figure}

Next, the assumption of stationarity and weak dependence is investigated.
In Fig.~\ref{fig:app_ACS}, the robustly estimated (by substituting the median for the mean) normalized auto-correlation sequences (ACSs) of the measurement data, pre-processed as in \eqref{app:prenormshift}, and of the same data split into 4 consecutive non-overlapping windows are plotted.
The data appears to be wide-sense (second-order) stationary, as the 4 ACSs in red are all similar.
Furthermore, the data is weakly dependent (required for the validity of the asymptotic distributions), as the ACSs decay to zero after approximately 20 samples.

\subsection{Reference methods}\label{app:weak_ref}
For the two reference methods a whitening transformation \cite{kessy2018optimal} is applied as a first step, for which knowledge of the covariance matrix is required.
As explained above, a 10-fold cross-validation procedure is applied (no validation set is needed, as we have no hyperparameters).
The covariance matrix is estimated from the training set, and the detectors' performances are evaluated on the test set.
To obtain a whitening matrix, we make use of \eqref{equ:app_asympt_cov_inv} combined with a robust estimate for the power spectral density.
This robust estimate averages multiple Schuster periodograms using the median instead of the mean, an established method in signal processing \cite{martin1982robust,zoubir2012robust}.
The whitening matrix for the reference methods is thus given by
\begin{equation}
    \sqrt{\boldsymbol\Sigma^{-1}}\approx\sum\limits_{n=0}^{N-1}\frac{\mathbf{v}_i\mathbf{v}_i^H}{\sqrt{P_{xx}(f_i)}},
\end{equation}
where $P_{xx}(f_i)$ is replaced by its robust estimate.
With this, the data is transformed as $\mathbf{y}=g(\sqrt{\boldsymbol\Sigma^{-1}}\mathbf{x})$, where $g(\cdot)$ is a function that is applied element-wise to its vector argument.
For the ``limiter function'' method $g(\cdot)$ is given by
\begin{equation}
    g(z)=
    \begin{cases}
        3 & z>3 \\
        z & -3<z<3\\
        -3 & z<-3
    \end{cases},
\end{equation}
for the ``sign function'' method $g(\cdot)$ is given by
\begin{equation}
    g(z)=
    \begin{cases}
        1 & z>0\\
        0 & z=0\\
        -1 & z<0
    \end{cases},
\end{equation}
and for the ``no nonlinearity'' method, $g(\cdot)$ is the identity function $g(z)=z$.
Afterwards, the GLR statistic for linear parameters with observation matrix $\mathbf{D}=\sqrt{\boldsymbol\Sigma^{-1}}\mathbf{H}$ in white Gaussian noise \cite[p.274]{kay1998fundamentals}, given by
\begin{equation}
    \mathbf{y}^\top \mathbf{D}(\mathbf{D}^\top \mathbf{D})^{-1}\mathbf{D}^\top \mathbf{y},
    \label{equ:app_GLRT_Gauss}
\end{equation}
is computed and compared to a threshold.
Note that for IID non-Gaussian noise, the asymptotically optimal Rao detector can be obtained by applying the GLR statistic to the score-transformed data.

\subsection{Neural network training and evaluation}\label{app:weak_neural}
In the following, we cover the details of neural network training and evaluation, which was implemented in Python using the deep learning library PyTorch \cite{paszke2019pytorch}.

As described in SI~\ref{app:weak_magnetic}, a nested cross-validation procedure is employed.
In tab.~\ref{tab:app_hyper}, the hyperparameters, some of which are optimized via grid search on the validation set, are listed.
The rather specific task of learning the locally optimal nonlinearity for weak signal detection enables us to impose reasonable constraints on most of the hyperparameters.
Additionally, as there is only limited data available, setting constraints helps to largely avoid overfitting.
\begin{table}[]
\begin{tabular}{|l|l|}
\hline
\textbf{parameter}           & \textbf{value / method} \\
\hline
number of conv1D layers      & 3 \\
number of channels           & 20\\
filter size                  & 3 \\
optimizer                    & AdamW \\
step size $\Delta\theta$ of difference quotient & 1e-2 \\
patience early stopping & 3 epochs \\
batch size & data set size \\
learning rate                & 1e-4 \\
L2 regularization weight & 1e-5\\
\hline
number of epochs & early stopping \\
\hline
\end{tabular}
\caption{The fixed (top) and variable (bottom) hyperparameters. The number of epochs is optimized on the validation set (early stopping).}
\label{tab:app_hyper}
\end{table}

\begin{figure}
    \centering
    \includegraphics[width=0.8\linewidth]{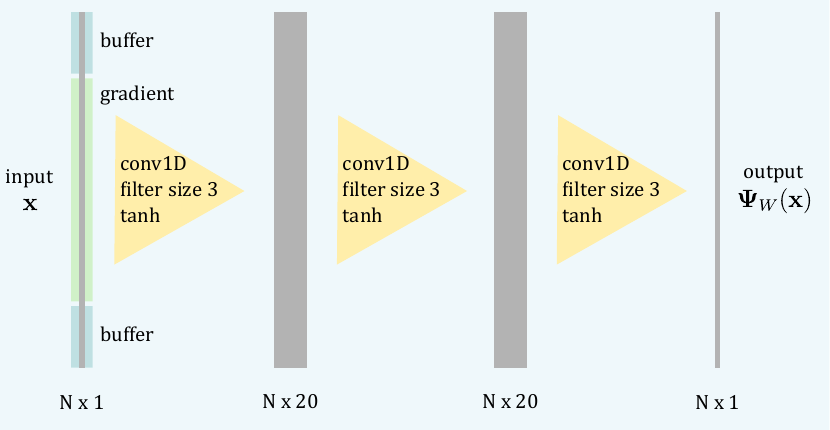}
    \caption{Schematic drawing of the convolutional neural network architecture. The vertical grey bars depict the layers of the neural network, and the yellow triangles indicate the filtering operation and the application of the (tanh) activation function. During training, only the samples of the shortened input sequence indicated in green are shifted to mitigate edge effects when calculating the gradient.}
    \label{fig:app_CNN}
\end{figure}

\begin{figure}
    \centering
    \includegraphics[width=0.5\linewidth]{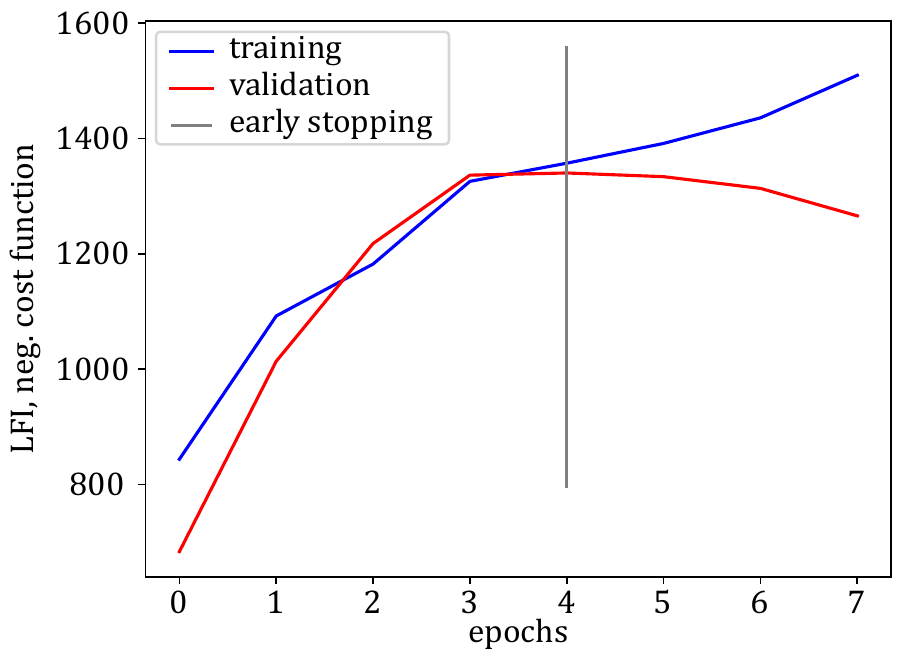}
    \caption{Exemplary illustration of early stopping to limit the effect of overfitting on the training data and to find the number of epochs the network is trained. Even though the LFI (and its trace) are theoretically bounded by the FI, overfitting can occur, since we are optimizing only an estimate of the asymptotic trace of the LFI.}
    \label{fig:app_early_stopping}
\end{figure}

In Fig.~\ref{fig:app_CNN}, a sketch of the architecture of our small convolutional neural network is depicted.
Inspired by the shape of typical nonlinearities for weak signal detection \cite{song2002advanced,kay1998fundamentals}, this architecture resembles a limiter function that reduces the effect of outliers in the data and is also capable of entangling complicated temporal dependencies in the input data.
Under the assumption of a symmetric noise distribution, the nonlinearity is an odd function.
We enforce this by using tanh activations and omitting the bias parameters in the conv1D layers.
The overall network size is chosen to be very small, based on the assumption that the noise distribution is weakly dependent and stationary, which should result in an optimal nonlinearity of limited complexity, as discussed in SI~\ref{app:weak_magnetic}.
Both the filter size and the number of layers are chosen as 3.
As shown in Fig.~\ref{fig:app_early_stopping}, the small model is capable of maximizing the LFI of the training data beyond that of the validation data, indicating that overfitting occurs (if it were not for the early stopping).
Thus, larger models are not necessary for the problem at hand.

If the amount of training data available is limited, designing a small, efficient network is crucial.
The cost function \eqref{equ_cost_function} is of asymptotic nature, therefore, for a finite sequence length $N$, it is only an approximation for the (negative) trace of the LFI (even before LFI is replaced by the estimated LFI).
If the architecture of the neural network is too large, the network might learn to decrease the cost function indefinitely, without actually increasing the LFI (if it were not for the early stopping).
This potential instability is caused not only by model complexity, but also by the level of noise in the cost function, as discussed in the following.

Probably the most important single hyperparameter is the step size $\Delta\theta$ that is used to numerically evaluate the gradient of the mean via a central difference quotient.
If its value is too low, the estimate of the trace of the LFI is too noisy, and the training procedure tends to be unstable.
If its value is too high, the approximation of the gradient via a central difference quotient is too crude, and we are essentially optimizing the coarse LFI \cite{kohn2016correlations}.
Nevertheless, it may be preferable to choose a step size that is slightly too high rather than too low, to avoid potential instabilities, even if this means deviating somewhat from the optimal local nonlinearity. 
A manual parameter search led to a value of $\Delta\theta=0.01$, a tradeoff between stability and validity of the central difference quotient.
Note that $\Delta\theta$ is highly dependent on the specifics of the problem at hand.
For large amounts of training data, a wider range of step sizes might lead to good results.

For the learning rate, we chose a value of $10^{-4}$ that leads to a reliable training procedure.
By reliable, we mean that the negative cost on the validation set has a clear maximum, as depicted exemplarily in Fig.~\ref{fig:app_early_stopping}.

In general, overfitting and instabilities are less of a problem as the size of the training data set increases and the sequence length $N$ grows.
Then, since we do not rely as much on the inductive bias anymore, a wider range of architectures can be considered.
We highlight that our approach is effective even with as few as $\lfloor10000/128\rfloor=78$ data sequences, which are split into training, testing, and validation sets.
Therefore, real-time applications featuring online network adaptation appear to be a promising direction for further development. 
This is also because the number of backpropagation steps required is quite low, see Fig.~\ref{fig:app_early_stopping} (one epoch equals one backpropagation step because the batch size equals the data set size).
\newpage
\bibliographystyle{unsrt}
\bibliography{sample}
\end{document}